\documentclass[journal=jacsat,manuscript=article]{achemso}

\usepackage{chemformula} 
\usepackage[T1]{fontenc} 
\usepackage{graphicx}
\usepackage{amsmath,amssymb,physics,dsfont}
\usepackage{color}
\usepackage{hyperref}
\usepackage{microtype}
\usepackage{braket}
\usepackage{defs-private}
\usepackage{qcircuit}
\usepackage{tikz}

\SectionNumbersOn
\author{Martin Mootz}
\email{mootz@iastate.edu}
\affiliation{Ames National Laboratory, U.S. Department of Energy, Ames, Iowa 50011, USA}

\author{Thomas Iadecola}
\affiliation{Ames National Laboratory, U.S. Department of Energy, Ames, Iowa 50011, USA}
\alsoaffiliation{Department of Physics and Astronomy, Iowa State University, Ames, Iowa 50011, USA}

\author{Yong-Xin Yao}
\email{ykent@iastate.edu}
\affiliation{Ames National Laboratory, U.S. Department of Energy, Ames, Iowa 50011, USA}
\alsoaffiliation{Department of Physics and Astronomy, Iowa State University, Ames, Iowa 50011, USA}

\title{Adaptive variational quantum computing approaches for Green's functions and nonlinear susceptibilities}

\begin{document}

\begin{abstract}
We present and benchmark quantum computing approaches for calculating real-time single-particle Green’s functions and nonlinear susceptibilities of Hamiltonian systems. The approaches leverage adaptive variational quantum algorithms for state preparation and propagation. Using automatically generated compact circuits, the dynamical evolution is performed over sufficiently long times to achieve adequate frequency resolution of the response functions. We showcase accurate Green's function calculations using a statevector simulator on classical hardware for Fermi-Hubbard chains of 4 and 6 sites, with maximal ansatz circuit depths of 65 and 424 layers, respectively, and for the molecule LiH with a maximal ansatz circuit depth of 81 layers. Additionally, we consider an antiferromagnetic quantum spin-1 model that incorporates the Dzyaloshinskii-Moriya interaction to illustrate calculations of the third-order nonlinear susceptibilities, which can be measured in two-dimensional coherent spectroscopy experiments. These results demonstrate that real-time approaches using adaptive parameterized circuits to evaluate linear and nonlinear response functions can be feasible with near-term quantum processors.
\end{abstract}

\section{Introduction}

In the pursuit of understanding and predicting the properties of quantum systems, the development of efficient computational methodologies has become imperative. Among the plethora of tools available, the calculation of Green's functions~\cite{bruus2004many,stefanucci2013nonequilibrium} and related high-order nonlinear susceptibilities~\cite{mukamel1995principles,Wan2019,Nandkishore2021} provide crucial insights into dynamical responses, correlation effects, and excitations within complex quantum materials and molecular systems. Specifically, many-particle Green's functions describe the correlations and interactions among particles in quantum systems via correlation functions or response functions and also facilitate the development of advanced theoretical frameworks for describing complex quantum phenomena such as the GW method~\cite{GWmethod} and dynamical mean-field theory~\cite{dmft_georges96, dmft_kotliar06, RMPNonequilibriumDMFT}. These methods rely heavily on Green's functions to investigate the rich phenomena exhibited by complex materials, including superconductivity~\cite{kopnin2001theory}, magnetism~\cite{schollwock2008quantum}, and topological phases~\cite{Gurarie2011}. Central to the analysis of Green's functions is the concept of spectral functions, which provide a direct link between theoretical calculations and measurable quantities in experiments, such as photoemission spectra~\cite{Damascelli_2004}, optical conductivity~\cite{mahan2012many}, or neutron scattering cross-sections~\cite{lovesey1986theory}. In contrast, nonlinear susceptibilities~\cite{mukamel1995principles,Wan2019,Nandkishore2021} offer a systematic approach to analyze the nonlinear optical response of quantum systems to multiple interacting laser pulses. Specifically, two-dimensional coherent spectroscopy (2DCS)~\cite{Kuehn2009, Kuehn2011, Junginger2012, woernerUltrafastTwodimensionalTerahertz2013, Maag2016,Yang2018TerahertzQT, Johnson2019, Higgs_2dTHz,Luo2019,Yang2019,Vaswani2020,Vaswani2020b,Yang2020lLightCS, mahmoodObservationMarginalFermi2021,Vaswani2021,Song2023UltrafastMP}  directly probes nonlinear susceptibilities by measuring the time-dependent coherent response to two laser pulses. 

Traditional methods for computing Green's functions and nonlinear susceptibilities, such as exact diagonalization~\cite{EDLib18}, density-matrix renormalization group~\cite{Schollwoeck2011TheDR}, or quantum Monte Carlo simulation~\cite{gubernatis_kawashima_werner_2016}, often face challenges in simulating large-scale systems at low temperatures. Nevertheless, in recent years the advent of quantum computing~\cite{feynman82qc} has opened up new avenues for tackling these challenges. These simulations hold promise in providing valuable insights into the properties of complex quantum materials that go beyond what is possible with classical computers~\cite{Daley2022}. So far, these simulations have been conducted on noisy intermediate-scale quantum devices~\cite{nisq, Hempel2018, hardware_efficient_vqe, Kemper_magnon, Chen2022}, which are restricted by the available number of qubits and by hardware noise. Several methods have been proposed to calculate Green’s functions and multi-time correlation functions~\cite{Pedernales2014} in frequency space using quantum-phase estimation~\cite{Baker2021,TroyerQCMB,Kosugi2020,Roggero2019}, and in the time domain using the Suzuki-Trotter decomposition of the time evolution operator~\cite{hybrd_dmft,Kreula2016,Chiesa2019,Kemper2024}. However, these techniques generally demand deep quantum circuits and large numbers of controlled operations, making them impractical for noisy intermediate-scale quantum hardware. To address these issues, several quantum circuit compression algorithms have been introduced. These algorithms calculate Green's functions in the frequency domain using the variational quantum eigensolver~\cite{endo2020calculation,Chen2021}, or in the time domain by simplifying the time evolution unitary operation using the coupled cluster Green’s function method~\cite{Keen2022hybridquantum}, Cartan decomposition~\cite{Kemper2023}, or variational quantum dynamics simulation algorithms~\cite{theory_vqs,endo2020calculation,Libbi2022}. The latter involves the preparation of a variational ansatz state to approximate the exact time-evolved state of the system. The equation of motion governing the time evolution of the variational parameters is derived based on the McLachlan variational principle~\cite{theory_vqs, Endo20variational, nagano2023quench}, which aims to minimize the distance between the variational state and the exact time-evolved state. 

Nonetheless, the efficacy of variational quantum dynamics simulations crucially depends on the flexibility of the variational ansatz to faithfully represent the dynamical states of the system. Using Hamiltonian variational ansatz (HVA)~\cite{endo2020calculation,Libbi2022}, the accuracy of the real-time Green’s function can be improved by increasing the number of layers, i.e., the depth of the ansatz. However, a large number of layers can be required to precisely describe the quantum state dynamics over sufficiently long times to achieve adequate resolution of correlation function in frequency space, leading to large circuit depths. Adaptive variational algorithms, such as the adaptive variational quantum dynamics simulation (AVQDS)~\cite{AVQDS} approach, can automatically generate problem-specific ans\"atze with reduced complexity compared to general problem-agnostic fixed ans\"atze~\cite{AVQDS, AVQITE, grimsleyAdaptiveVariationalAlgorithm2019}. In AVQDS, the McLachlan distance, a metric to measure the difference between the variational and exact state evolutions, is maintained below a predefined threshold throughout the time evolution by adaptively adding new parameterized unitaries chosen from a predetermined operator pool to the variational ansatz. This method has been applied in ref~\citenum{Gomes2023} to calculate the single-particle Green’s function using a Hadamard-test circuit to compute state overlaps. The result in ref~\citenum{Gomes2023} demonstrates that less quantum resources are required compared to variational quantum dynamics simulations with HVA, even as the accuracy over long simulation times is improved. However, the Hadamard-test circuit involves controlled multi-qubit rotation gates for state propagation, which can substantially increase the circuit complexity.

In this work, we employ the AVQDS algorithm to calculate single-particle Green’s functions and nonlinear susceptibilities along the real-time axis. 
In contrast to the controlled-unitaries-required (CUR) approach for overlap testing~\cite{Gomes2023}, we adopt the method presented in ref~\citenum{endo2020calculation}, which is controlled-unitaries-liberated (CUL) and applies to generic reference states besides the ground state. Instead, the method requires the parametrized circuit to directly simulate the dynamics of two quantum states, as opposed to one in the CUR method~\cite{Libbi2022,Gomes2023}. The CUL calculation is conveniently performed on circuits with an ancilla qubit plus a few controlled Pauli gates to mix the two quantum states, followed by state evolution. 
Although the simultaneous propagation of two states instead of one demands more flexible circuits, it allows one to leverage the variational degrees of freedom already existing in the initial parameterized state (e.g., the ground state) for variational state propagation using more compact circuits. We apply the CUL approach to calculate the single-particle Green's functions of fermionic models including molecules using the AVQDS approach instead of variational quantum dynamics simulations with HVA as in ref~\citenum{endo2020calculation}. Additionally, we extend this method to calculate nonlinear susceptibilities of quantum spin models, which depend on two times and thus requires double application of the AVQDS approach to evolve the quantum states. The presented AVQDS simulations are performed using statevector simulators on classical hardware to validate our methods using well-established classical approaches. These simulations also allow us to estimate the quantum resources required for specific applications. Such estimates, typically characterized by metrics like the number of CNOT gates in a circuit for near-term quantum computing, provide critical insights before transitioning to actual quantum hardware

To illustrate the calculation of Green's functions with the AVQDS approach, we study the single-particle Green’s functions of Fermi-Hubbard chains with $N=4$ and $N=6$ sites, corresponding respectively to 8 and 12 qubits. Additionally, we determine the Green's function of the molecule LiH where the encoding of the  Hamiltonian requires 10 qubits.  Using statevector simulations, we calculate the Green’s function dynamics in momentum space and discuss the required quantum resources for near-term applications measured by number of CNOT gates and circuit layers. Additionally, we compute the spectral function and compare the results obtained with the circuit simulator to exact results derived from the Lehmann representation of the Green’s function. To demonstrate the validity of our method to calculate high-order susceptibilities measured in two-dimensional coherent spectroscopy experiments, we consider an antiferromagnetic quantum high-spin model that incorporates the Heisenberg exchange and Dzyaloshinskii-Moriya interactions~\cite{mootz2023twodimensional}. Specifically, we calculate the third-order nonlinear susceptibility in the two-dimensional (2D) time and frequency domain for a two-site spin-1 model and compare with exact numerical results.

The paper is organized as follows. The AVQDS algorithm is briefly discussed in section~\ref{sec:avqds}. In section~\ref{sec:methods} we present the algorithm for calculating single-particle Green’s functions and the method for simulating nonlinear susceptibilities. Section~\ref{sec:apps} presents various applications of these methods, demonstrating their utility in practical scenarios. Specifically, in section~\ref{sec:GF_res} we calculate the Green’s function and corresponding spectral functions for $N=4$ and $N=6$ Fermi-Hubbard chains and benchmark the performance of the AVQDS approach. In section~\ref{sec:molecule}, we present the Green's function calculation for the molecule LiH. In section~\ref{sec:chi3}, we demonstrate the algorithm for simulating nonlinear susceptibilities by calculating the third-order susceptibility of a two-site spin-1 model. Finally, we conclude in section~\ref{sec:con} with a summary of findings and an outlook.

\section{AVQDS algorithm}
\label{sec:avqds}
In this work, we employ the AVQDS algorithm as a quantum computing approach to calculate Green's functions and high-order nonlinear susceptibilities. Below, we briefly discuss the key points of AVQDS, while a more detailed discussion can be found in refs~\citenum{AVQDS,mootz2023twodimensional}. The algorithm can be naturally extended from real-time dynamics to the adaptive variational quantum imaginary-time evolution (AVQITE) approach~\cite{AVQITE}, which is adopted for ground state preparation in the following calculations.
We begin by considering a quantum system initially in the pure state $|\Psi\rangle$, whose quantum dynamics is governed by a generally time-independent Hamiltonian $\h$. The dynamics of the system's density matrix $\rho=|\Psi\rangle\langle\Psi|$ is then determined by the von Neumann equation:
\begin{align}
	\frac{\mathrm{d}\rho}{\mathrm{d}t}=-\mathrm{i}\left[\hat{\mathcal{H}},\rho\right]\,.
\end{align}
In variational quantum dynamics simulations, the state $|\Psi\rangle$ is parameterized as $|\Psi[\bth]\rangle$, with $\bth(t)$ being a real-valued time-dependent variational parameter vector of dimension $N_\theta$~\cite{theory_vqs}. The evolution of $\bth$ is determined by equations of motion derived from the McLachlan variational principle~\cite{mclachlan64variational}, which minimizes the squared McLachlan distance $\mathcal{L}^2$ as the Frobenius norm ($\| \mathcal{O} \| \equiv \Tr[\mathcal{O}^\dag \mathcal{O}]$) between exact and variational evolving states:
\begin{align}
\label{eq:L2}
    \mathcal{L}^2&\equiv\bigg\|\sum_\mu\frac{\partial\rho[\bth]}{\partial\theta_\mu}\dot{\theta}_\mu+\mathrm{i}\left[\hat{\mathcal{H}},\rho\right]\bigg\|^2 \nonumber \\
    &=\sum_{\mu\nu}M_{\mu\nu}\dot{\theta}_\mu\dot{\theta}_\nu-2\sum_\mu V_\mu\dot{\theta}_\mu+2\,\mathrm{var}_{\bth}[\h]\,,
\end{align}
where the $N_\theta\times N_\theta$ matrix  $M$ and vector $V$ of dimension $N_\theta$ are defined as
\begin{align}
\label{eq:M}
M_{\mu,\nu}\equiv &\,\mathrm{Tr}\left[\frac{\partial\rho[\bth]}{\partial\theta_\mu}\frac{\partial\rho[\bth]}{\partial\theta_\nu}\right]=2\,\mathrm{Re}\left[\frac{\partial\langle\Psi[\bth]|}{\partial \theta_\mu}\frac{\partial|\Psi[\bth]\rangle}{\partial \theta_\nu}\right.\nonumber \\ &\left.+\frac{\partial\langle\Psi[\bth]|}{\partial \theta_\mu}|\Psi[\bth]\rangle\frac{\partial\langle\Psi[\bth]|}{\partial \theta_\nu}|\Psi[\bth]\rangle\right]\,, \\
	V_\mu=&\,2\,\mathrm{Im}\left[\frac{\partial\langle\Psi[\bth]|}{\partial \theta_\mu}\h|\Psi[\bth]\rangle+\langle\Psi[\bth]|\frac{\partial|\Psi[\bth]\rangle}{\partial \theta_\mu}\langle \h\rangle_{\bth}\right]\,,
 \label{eq:V}
\end{align}
with $\langle \h\rangle_{\bth}=\langle\Psi[\bth]|\h|\Psi[\bth]\rangle$. The real symmetric matrix $M$ is directly related to the quantum Fisher information matrix~\cite{meyer2021fisher}, with the second term within the bracket accounting for the global phase contribution~\cite{theory_vqs}. In the last term of eq~\ref{eq:L2}, $\mathrm{var}_{\bth}[\h]=\langle\h^2\rangle_{\bth}-\langle\h\rangle^2_{\bth}$ is the variance of $\h$ in the variational state $|\Psi[\bth]\rangle$.   
The minimization of $\mathcal{L}^2$ with respect to $\{\dot{\theta}_\mu\}$ leads to the equation:
\begin{align}
\label{eq:theta_eom}
    \sum_\nu M_{\mu\nu}\dot{\theta}_\nu=V_\mu\,,
\end{align}
which governs the dynamics of the variational parameters. The optimized McLachlan distance for the variational ansatz $|\Psi[\bth]\rangle$,
\begin{align}
    L^2=2\,\mathrm{var}_{\bth}[\h]-\sum_{\mu}V_\mu \dot{\theta}_\mu\,,
\end{align}
is a metric quantifying the accuracy of the variational dynamics.

In the AVQDS framework, the ansatz takes a pseudo-Trotter form:
\begin{align}
\label{eq:ansatz}
|\Psi[\bth]\rangle=\prod_{\mu=1}^{N_\theta}\mathrm{e}^{-\mathrm{i}\theta_\mu\hat{\mathcal{A}}_\mu}|\varphi_0\rangle\,.
\end{align}
Here, $\hat{\mathcal{A}}_\mu$ ($\mu=1,\cdots,N_{\theta})$ represent Hermitian generators, while $\ket{\varphi_0}$ denotes the reference state. To maintain the McLachlan distance $L^2$ below a given threshold $L^2_\mathrm{cut}$ throughout the time evolution, operators are dynamically selected from a predefined pool of generators, which expands the set of unitaries in the ansatz eq~\ref{eq:ansatz}. The selection criterion is to add to the ansatz the operator that maximally reduces $L^2$, so as to achieve accurate dynamics simulations with the minimal number of unitaries. In the initial proposal of the AVQDS approach~\cite{AVQDS}, operators were chosen one by one; here we adopt a modified strategy. Specifically, multiple operators, rather than a single one, are simultaneously chosen according to the $L^2$-criterion at each iteration, subject to the constraint that these operators act on disjoint sets of qubits. This approach takes the spatial compactness of the circuits into account, and significantly reduces the circuits depth with only a small change in the total number of CNOT gates, 
The ansatz expansion is done by first calculating the McLachlan distance $L^2_{\nu}$ for a new variational ansatz of the form $e^{-i\theta_\nu\hat{\mathcal{A}}_\nu}|\Psi[\bth]\rangle$ for all generators $\hat{\mathcal{A}}_\nu$ from a predefined operator pool of size $N_\mathrm{p}$. The resulting $L^2_{\nu}$ are ordered in ascending order. In the first step, the operator $\hat{\mathcal{A}}_\nu$ leading to the smallest $L^2_\nu$ is added to the ansatz eq~\ref{eq:ansatz}, which increases $N_\theta \rightarrow N_\theta + 1$ in eq~\ref{eq:ansatz}. In the next step, the operator with the next smallest $L^2_\nu$ is appended, provided it acts on different qubits than the first operator. This process continues by adding operators with successively larger or equal $L^2_\nu$ that have disjoint supports from all previously appended operators during this iteration, until all qubits are covered or no more suitable operators are found. Note that this method does not introduce additional $L^2_\nu$ measurements, as $L^2$ for each pool operator is measured once per iteration. The variational parameters $\theta_\nu$ are set to zero for all operators appended to the ansatz eq~\ref{eq:ansatz} during the iteration step. This does not alter the ansatz state but can modify the McLachlan distance $L^2$ due to a non-zero derivative with respect to $\theta_\nu$. Finally, $L^2$ is calculated for the new ansatz and compared with $L^2_\mathrm{cut}$. This adaptive procedure is repeated until the McLachlan distance $L^2$ of the new ansatz falls below $L^2_\mathrm{cut}$.

Following the adaptive ansatz expansion procedure, the variational parameters are evolved in time according to eq~\ref{eq:theta_eom}, which corresponds to a system of linear ordinary differential equations. In the simulations performed in this work, we solve eq~\ref{eq:theta_eom} using a fourth-order Runge-Kutta method. This method provides a higher accuracy with a truncation error of order $(\delta t)^4$, compared to the Euler method with a truncation error of order $(\delta t)^2$.  While this method involves computational overhead due to each time step requiring four micro-time steps to update the variational parameters (compared to Euler's method, which updates parameters in a single step) it still substantially reduces the total number of time steps and circuit complexity, as demonstrated in ref~\citenum{mootz2023twodimensional}. Additionally, the time step $\delta t$ in the AVQDS approach is dynamically adjusted to ensure that $\max_{0\leq\mu<N_\theta}|\delta\theta_\mu|$ remains below a predefined maximal step size $\delta\theta_\mathrm{max}$, where $\delta\theta_\mu=\theta_\mu(t+\delta t)-\theta_\mu(t)$. In the simulations we set $\delta\theta_\mathrm{max}=0.01$. Furthermore, to mitigate potential numerical issues associated with inverting the matrix $M$ in eq~\ref{eq:theta_eom}, we employ the Tikhonov regularization approach, where a small number of $\delta=10^{-6}$ is added to the diagonal of $M$. This stabilizes the matrix inversion process, especially when $M$ possesses a high condition number.

To estimate the cost of measuring $M$, $V$, $\langle\h\rangle_\theta$, and $\langle\h^2\rangle_\theta$ on quantum hardware, we assume that the wavefunction ansatz has $N_\theta$ variational parameters. In addition, we assume that the Hamiltonian consists of $N_\mathrm{H}$ Pauli strings, which also corresponds to the number of generators in the operator pool when the Hamiltonian operator pool is chosen, as is the case in the Green's function calculation for the Fermi-Hubbard chains and the LiH molecule in sections~\ref{sec:GF_res} and \ref{sec:molecule}. As shown in ref~\citenum{AVQDS}, the upper bound on the number of distinct direct measurement circuits and generalized Hadamard test circuits is then given by $(N_\mathrm{H} + 2) N_\theta + N_\mathrm{H} + N_\mathrm{H}^2$ and $N_\mathrm{H}(N_\theta - 1) + N_\theta(N_\theta - 1)/2$, respectively. The adaptive ansatz expansion
procedure requires an additional $N_\mathrm{H}(N_\theta - 1)$ generalized Hadamard test circuits. For example, for the $N=4$-site ($N=6$-site) Fermi-Hubbard chain, the Hamiltonian consists of $N_\mathrm{H}=16$ ($N_\mathrm{H}=26$) terms while the maximum $N_\theta$ among all $I^{p,q}_{r, s}$ at time $t=10$ is $N_\theta=221$ ($N_\theta=2021$). In this situation, a single Runge-Kutta time step requires $1.3\times 10^5$ ($8.8\times 10^6$) measurements for $N=4$ ($N=6$) while the adaptive ansatz expansion procedure demands an additional $3.5\times 10^3$ ($5.2\times 10^4$) measurements for $N=4$ ($N=6$). Note that each time step using the Runge-Kutta method involves four measurements of the matrix $M$, vector $V$, $\langle\h\rangle_\theta$, and $\langle\h^2\rangle_\theta$.

Regarding the sensitivity to noise of the AVQDS approach, we have studied the impact of noise on the AVQITE approach for ground state preparation in ref~\citenum{getelina2023adaptive}. Specifically, we investigated how sampling noise---coherent errors resulting from imperfect gate operations, and stochastic errors caused by qubit decoherence, dephasing, and relaxation---affect the AVQITE simulations. We considered a fixed uniform single-qubit gate error rate of $10^{-4}$, which corresponds to the value realized in current hardware, and a uniform two-qubit error rate 
in the range $[10^{-4}, 10^{-2}]$. Additionally, we used $2^{14}$ shots for each measurement circuit. We found that, for a two-qubit noise level of $10^{-2}$ relevant for current hardware, the ground state energy error is about $12\%$. This decreases to $3\%$ for a two-qubit gate error rate of $10^{-3}$. We expect comparable errors for the dynamics simulations using AVQDS.

To benchmark the performance of AVQDS, we compare the AVQDS results with results utilizing exact diagonalization for state propagation:
\begin{align}
|\Psi[t+\delta t]\rangle = \mathrm{e}^{-\mathrm{i}\delta t\h}|\Psi[t]\rangle\,.
\label{eq:ED}
\end{align}
This simulation is performed on a uniformly discretized time grid with a step size of $\delta t = 0.002$. Since our study focuses on time-independent Hamiltonians, the error due to finite time mesh here is zero.

\section{Methods \label{sec:methods}}

\subsection{\label{sec:GF} Computation of Green's functions with AVQDS}


We begin by considering a time-independent fermionic Hamiltonian $\h$, expressed in terms of fermionic operators $\cc_p$ and $\ca_p$ that create and annihilate an electron in orbital site $p=(j,\sigma)$. Here $j$ is the site index and $\sigma$ corresponds to the spin of the fermion. The dynamics of the system at zero temperature is described by the retarded single-particle Green's function:
\begin{align}
\label{eq:Gr}
    G^\mathrm{R}_{p,q}(t)=-\mathrm{i}\Theta(t)&\left[\bra{\mathrm{G}}\ca_p(t)\cc_q(0)\ket{\mathrm{G}}\right. \nonumber \\ &\left.+\bra{\mathrm{G}}\cc_q(0)\ca_p(t)\ket{\mathrm{G}} \right]\,.
\end{align}
Here, $\ca_p(t)=\mathrm{e}^{\mathrm{i}\h t}\ca_p\mathrm{e}^{-\mathrm{i}\h t}$ corresponds to the Heisenberg representation of the fermionic operator $\ca_p$, $\Theta(t)$ is the Heaviside step function, and $\ket{\mathrm{G}}$ denotes the ground state of Hamiltonian $\h$.

\begin{figure}[t!]

\begin{tikzpicture}
        \node[anchor=south west, inner sep=0] (image) at (0,0){};
        \node[above left] at (-11.5, 0) {(a)};
\end{tikzpicture}

\vspace{-5pt}
\centerline{\Qcircuit @C=1.3em @R=1.3em{ \lstick{|0\rangle } & \gate{H} &\ctrl{1} &\qw & \gate{X} & \ctrl{1} & \gate{H} &\meter\qw\\ \lstick{|\varphi_0\rangle} & \gate{U_\mathrm{G}}&\gate{P_\beta} & \gate{\mathrm{e}^{-\mathrm{i}\h t}} &\qw & \gate{P_\alpha} & \qw & \qw 
\gategroup{1}{2}{2}{4}{0.7em}{--}
}}

\begin{tikzpicture}
        \node[anchor=south west, inner sep=0] (image) at (0,0){};
        \node[above left] at (-11.5, 0) {(b)};
\end{tikzpicture}
    
\vspace{-5pt}
\centerline{\Qcircuit @C=1.0em @R=1.3em{ \lstick{|0\rangle } & \gate{H} &\ctrl{1} &\qw & \gate{X} & \ctrl{1} & \gate{H} &\meter\qw\\ \lstick{|\varphi_0\rangle} & \gate{U_\mathrm{G}[\bth^1]}&\gate{P_\beta} & \gate{U_t[\bth^2]} &\qw & \gate{P_\alpha} & \qw & \qw 
\gategroup{1}{2}{2}{4}{0.7em}{--}
}}

\caption{Hadamard test circuit to compute the Green's function.
(a) Circuit to measure the Green's function component $I_{\alpha,\beta}^{p,q}$, eq~\ref{eq:Ipq}, using the exact time-evolution operator $\mathrm{e}^{-\mathrm{i}\h t}$. 
The ancillary qubit, initially in the state $\ket{0}$, is represented by the upper horizontal line. $X$ and $H$ denote Pauli-$X$ and Hadamard gates on the ancilla, respectively. A register of qubits for the physical system of interest, initially in a reference product state $\ket{\varphi_0}$, is denoted by the lower horizontal line. The application of the general multi-qubit Pauli gates $P_\alpha$ and $P_\beta$ is controlled by the ancilla qubit, while the unitary operator $U_\mathrm{G}$ prepares the ground state, $\ket{\mathrm{G}}=U_\mathrm{G}\ket{\varphi_0}$. 
(b) Measurement circuit for $I_{\alpha,\beta}^{p,q}$ using a variational state evolution circuit. The state propagation circuit highlighted with the dashed rectangle in (a) is replaced by the (adaptive) variational circuit in (b), where the angles in the parameterized unitaries $U_\mathrm{G}[\bth^1]$ and $U_t[\bth^2]$ evolve with time.
The results are obtained from $Z$-basis measurements on the ancilla qubit.}
\label{fig1}    
\end{figure}
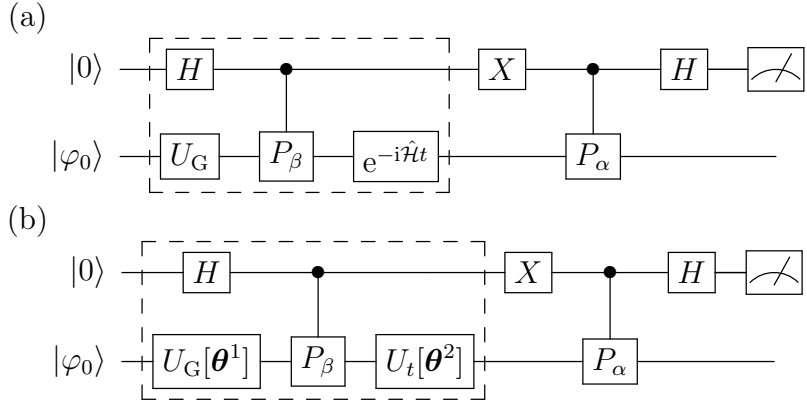

To calculate eq~\ref{eq:Gr} using a quantum computer, we adopt the Jordan-Wigner transformation~\cite{map_jw} to map the fermionic operators to Pauli operators as: 
\begin{align}
\label{eq:JW}
    \ca_p=\sum_{\alpha}\eta^{(p)}_\alpha P^{(p)}_\alpha\,.
\end{align}
Here, $\eta^{(p)}_\alpha$ are complex-valued numbers and $P^{(p)}_\alpha$ are Pauli words up to a weight of $N_q$, which is the number of qubits required to encode the fermionic Hamiltonian $\h$. Using transformation eq~\ref{eq:JW}, eq~\ref{eq:Gr} can be rewritten as:
\begin{align}
    G^\mathrm{R}_{p,q}(t)&=-i\,\Theta(t)\sum_{\alpha,\beta}\eta^{(p)}_\alpha\eta^{(q)}_\beta\left[\bra{\mathrm{G}}\mathrm{e}^{\mathrm{i}\h t}P^{(p)}_\alpha\mathrm{e}^{-\mathrm{i}\h t}P^{(q)}_\beta\ket{\mathrm{G}} \right. \nonumber \\
    & \qquad\qquad\qquad\qquad\;\;\left. +\bra{\mathrm{G}}P^{(q)}_\beta\mathrm{e}^{\mathrm{i}\h t}P^{(p)}_\alpha\mathrm{e}^{-\mathrm{i}\h t}\ket{\mathrm{G}} \right]\nonumber \\
    &\equiv-2\mathrm{i}\,\Theta(t)\sum_{\alpha,\beta}\eta^{(p)}_\alpha\eta^{(q)}_\beta I^{p,q}_{\alpha,\beta}\,,
    \label{eq:Gr_trafo}
\end{align}
with
\begin{align}
    I^{p,q}_{\alpha,\beta}\equiv \mathrm{Re}\left[\bra{\mathrm{G}}\mathrm{e}^{\mathrm{i}\h t}\P^{(p)}_\alpha\mathrm{e}^{-\mathrm{i}\h t}P^{(q)}_\beta\ket{\mathrm{G}}\right]\,.
    \label{eq:Ipq}
\end{align}
Equation~\ref{eq:Ipq} can be measured using a Hadamard test circuit on a quantum computer, as shown in Figure~\figref{fig1}{a}. The time evolution operator $\mathrm{e}^{-\mathrm{i}\h t}$ can be approximated using Trotter decomposition~\cite{lloyd1996, Trotter_dynamics_Knolle,childs2018quamtumsimulation} or variational quantum dynamics simulation algorithms~\cite{theory_vqs}, including AVQDS~\cite{AVQDS}. 

In this work, we first use the AVQITE algorithm to prepare the ground state of a Hamiltonian system, $\ket{G[\bth^1]}=U_\mathrm{G}[\bth^1]\ket{\varphi_0}$, with a series of parameterized unitaries $U_\mathrm{G}[\bth^1]$ applied to a reference product state $\ket{\varphi_0}$. The time propagation of the ancilla and physical register, as outlined by the dashed box in Figure~\figref{fig1}{a}, is achieved by a parameterized circuit highlighted by the dashed box in Figure~\figref{fig1}{b}. The additional unitaries $U_t[\bth^2]$ are automatically generated following the AVQDS algorithm. The rotation angles $\bth^1$ and $\bth^2$ both evolve according to the equation of motion eq~\ref{eq:theta_eom}. In other words, AVQDS is adopted for the time propagation of the state
\be
\ket{\Psi} = \frac{1}{\sqrt{2}}\ket{0}\otimes\ket{G}+\frac{1}{\sqrt{2}}\ket{1}\otimes P_\beta\ket{G} \label{eq:qc1}
\ee
under a Hamiltonian that acts only on the physical register. As shown in Appendix~\ref{sec:G_qc}, the circuit in Figure~\figref{fig1}{b} measures the Green's function component as
\be
I^{p,q}_{\alpha,\beta} \approx \mathrm{Re}\left[\bra{G[\bth^1]}U_t^\dagger[\bth^2]\P^{(p)}_\alpha U_t[\bth^2] P^{(q)}_\beta \ket{G[\bth^1]}\right]. \label{eq:Ipqv}
\ee

\subsection{\label{sec:chi} Computation of nonlinear susceptibilities with AVQDS}

\subsubsection*{Susceptibility expansion}

While full time-dependent non-equilibrium calculations are a powerful approach for capturing nonlinear effects, particularly when considering the entire spectrum of dynamics beyond weak signals, nonlinear susceptibilities offer meaningful and interpretable insights in scenarios where linear approximations fail but where a fully time-dependent non-equilibrium approach may not be necessary or practical. Nonlinear susceptibilities, in particular, provide a complementary perspective that is often more tractable and insightful, enabling the analysis of higher-order effects in a controlled manner. This approach makes it easier to isolate and understand specific nonlinear contributions and is also valuable for understanding the onset of nonlinear behavior before a full breakdown of the linear response.

Before we discuss the formalism, we begin with a brief discussion of the susceptibility expansion for nonlinear responses measured in 2DCS experiments. We consider an $N$-site quantum spin system described by spin-$s$ operators $\hat{S}^p_i$ ($p=x,y,z$) at site $i$. The transmitted magnetic field measured in 2DCS experiments on magnetic systems is determined by the magnetization $\mathbf{M}(t)\equiv \langle \hat{\mathbf{S}}^\mathrm{tot}\rangle=\langle\Psi[t]|\hat{\mathbf{S}}^\mathrm{tot}|\Psi[t]\rangle$. Here, $\hat{\mathbf{S}}^\mathrm{tot}=\sum_{j=1}^N \hat{\mathbf{S}}_j$ is the total spin operator and $|\Psi[t]\rangle$ corresponds to the quantum state of the system at time $t$.
In 2DCS the quantum spin system is excited by two magnetic field pulses polarized along the $\beta$- and $\gamma$-directions, separated in time by $\tau$. The total applied magnetic field can be written as $\mathbf{B}(t) = B^\beta_1(t)\boldsymbol{\beta} + B^\gamma_2(t - \tau)\boldsymbol{\gamma}$, with the two pulses centered at time $t=0$ and $t=\tau$, while $\boldsymbol{\beta}$ and $\boldsymbol{\gamma}$ denote the unit vectors along the $\beta$ and $\gamma$ directions. The differential transmitted magnetic field along the $\alpha$ direction, $B^\alpha_\mathrm{NL}$, as measured in 2DCS experiments, is proportional to the nonlinear differential magnetization:
\begin{align}
\label{eq:Mnl}
M^\alpha_\mathrm{NL}(t,\tau)=M^\alpha_\mathrm{12}(t,\tau)-M^\alpha_\mathrm{1}(t)-M^\alpha_\mathrm{2}(t,\tau)\,,    
\end{align} 
which depends on time $t$ and the inter-pulse delay $\tau$. Here,  $M^\alpha_{12}(t,\tau)$ denotes the magnetization dynamics induced by both pulses, while ${M}^\alpha_{1}(t)$ and ${M}^\alpha_{2}(t,\tau)$ denote the magnetization dynamics induced by pulses 1 and 2, respectively. As demonstrated in ref~\citenum{mootz2023twodimensional}, the 2DCS spectra obtained by a 2D Fourier transform of eq~\ref{eq:Mnl} can be interpreted by applying a susceptibility expansion~\cite{mukamel1995principles,Wan2019,Nandkishore2021} of $M^\alpha_\mathrm{NL}(t,\tau)$. By approximating the pulse shapes by $\delta$-functions, the applied magnetic field can be written as
\begin{align}
\mathbf{B}(t) = A_1^\beta\,\delta(t)\boldsymbol{\beta} + A_2^\gamma\,\delta(t-\tau)\boldsymbol{\gamma}\,, 
\end{align}	
where $A_i$ corresponds to the $i$th pulse area. As a result, the nonlinear magnetization density along the $\alpha$-direction at time $t+\tau$ can be expressed in terms of nonlinear susceptibilities $\chi^{(n)}$ of order $n$~\cite{Wan2019,Choi2020,Nandkishore2021}:
\begin{align}
	&M_\mathrm{NL}^\alpha(t+\tau)/N \nonumber \\ 
        &\equiv (M^\alpha_\mathrm{12}(t+\tau)-M^\alpha_\mathrm{1}(t+\tau)-M^\alpha_\mathrm{2}(t+\tau))/N \nonumber \\
	&= \chi^{(2)}_{\alpha\beta\gamma}(t,\tau)\,A^\beta_1 A^\gamma_2  \nonumber \\
	&+\chi^{(3)}_{\alpha\beta\gamma\delta}(t,\tau,0)\,(A^\beta_1)^2 A^\gamma_2 +\chi^{(3)}_{\alpha\beta\gamma\delta}(t,0,\tau)\,A^\beta_1 (A^\gamma_2)^2 \nonumber \\
	&+\mathcal{O}(B^4)\,.
 \label{eq:Mexp}
\end{align}
The second-order susceptibility in the above equation is explicitly given by~\cite{Wan2019,Choi2020}
\begin{align}
	&\chi^{(2)}_{\alpha\beta\gamma}(t,\tau)=\nonumber \\
 &-\frac{1}{N}\Theta(t)\Theta(\tau)\langle\left[\left[\hat{M}^\alpha(t+\tau),\hat{M}^\beta(\tau)\right],\hat{M}^\gamma(0)\right]\rangle\,,
\end{align}
while the third-order susceptibilities are defined by~\cite{Nandkishore2021}
\begin{align}
\label{eq:chi3}
	&\chi^{(3)}_{\alpha\beta\gamma\delta}(t,\tau,0)= \nonumber \\
 &-\frac{\mathrm{i}}{N}\Theta(t)\Theta(\tau)\langle\left[\left[\left[\hat{M}^\alpha(t+\tau),\hat{M}^\beta(\tau)\right],\hat{M}^\gamma(0)\right],\hat{M}^\delta(0)\right]\rangle\,, \\
	&\chi^{(3)}_{\alpha\beta\gamma\delta}(t,0,\tau)= \nonumber \\
 &-\frac{\mathrm{i}}{N}\Theta(t)\Theta(\tau)\langle\left[\left[\left[\hat{M}^\alpha(t+\tau),\hat{M}^\beta(\tau)\right],\hat{M}^\gamma(\tau)\right],\hat{M}^\delta(0)\right]\rangle\,.
\end{align}
In this paper, we demonstrate the calculation of the third-order susceptibility $\chi^{(3)}_{\alpha\beta\gamma\delta}(t,\tau,0)$ using a quantum computing approach. The second-order $\chi^{(2)}_{\alpha\beta\gamma}(t,\tau)$ and third-order $\chi^{(3)}_{\alpha\beta\gamma\delta}(t,0,\tau)$ as well as higher-order susceptibilities can be evaluated similarly. This approach is complementary to the method of direct observable dynamics simulations we developed earlier by disentangling contributions from quantum processes of different orders~\cite{mootz2023twodimensional}.

\subsubsection*{\label{sec:chi_form} Formalism}

We next discuss how $\chi^{(3)}_{\alpha\beta\gamma\delta}(t,\tau,0)$ can be measured on a quantum computer. By using $\hat{M}^\alpha=\sum_{j=1}^{N}\hat{S}^\alpha_j$, the third-order susceptibility eq~\ref{eq:chi3} can be written as:
\begin{align}
\label{eq:chi3b}
	&\chi^{(3)}_{\alpha\beta\gamma\delta}(t,\tau,0) \nonumber \\
 &=\frac{2}{N}\Theta(t)\Theta(\tau)\sum_{j,k,l,m=1}^{N}\mathrm{Im}\left[\langle \hat{S}^\alpha_j(t+\tau)\hat{S}^\beta_k(\tau)S^\gamma_l(0){S}^\delta_m(0)\rangle\right.\nonumber \\
&\qquad\qquad\qquad\qquad\quad\left. +\langle S^\delta_m(0){S}^\gamma_l(0)\hat{S}^\alpha_j(t+\tau)\hat{S}^\beta_k(\tau)\rangle \right.\nonumber \\
&\qquad\qquad\qquad\qquad\quad\left. -\langle\hat{S}^\gamma_l(0) \hat{S}^\alpha_j(t+\tau)\hat{S}^\beta_k(\tau)S^\delta_m(0)\rangle\right.\nonumber \\
&\qquad\qquad\qquad\qquad\quad\left. -\langle\hat{S}^\delta_l(0) \hat{S}^\alpha_j(t+\tau)\hat{S}^\beta_k(\tau)S^\gamma_m(0)\rangle\right]\,.
\end{align}
To evaluate eq~\ref{eq:chi3b} for a generic spin-$s$ model using a quantum computer, one can either use a qudit-based quantum device~\cite{wang2020,Ogunkoya2024QutritCA} or bosonic quantum devices~\cite{dutta2024simulating, stavenger2022c2qa} where the number of qudit levels/qumodes corresponds to the number of spin states $2s +1$, or map the spin-$s$ levels to qubits using transformations such as the Gray code or binary encoding~\cite{mootz2023twodimensional,getelina2024}. In this paper, we choose the latter approach as qubit-based platforms are currently the most widely available. The transformation of spin-$s$ operators to multi-qubit operators can be written as 
\begin{align}
    \hat{S}^\alpha_j=\sum_{p=1}^{n_\alpha} \eta^\alpha_{j,p}P^\alpha_{j,p}\,.
    \label{eq:s_trafo}
\end{align}
Here, the transformation contains $n_\alpha$ terms and the index $j=1,\dots,N$ labels the physical site; $\eta^\alpha_{j,p}$ are real-valued coefficients and $P^\alpha_{j,p}$ are Pauli words. The encoding of a spin-$s$ site requires $n_q$ qubits such that the system contains $N_q=n_q N$ qubits in total. Using $\hat{S}^\alpha_j(t)=\mathrm{e}^{\mathrm{i}\h t}\hat{S}^\alpha_j\mathrm{e}^{-\mathrm{i}\h t}$ and eq~\ref{eq:s_trafo}, eq~\ref{eq:chi3b} becomes
\begin{align}
\label{eq:chi3_trans}
	&\chi^{(3)}_{\alpha\beta\gamma\delta}(t,\tau,0) \nonumber \\
 &=\frac{2}{N}\Theta(t)\Theta(\tau)\sum_{j,k,l,m=1}^{N}\sum_{p=1}^{n_\alpha}\sum_{q=1}^{n_\beta}\sum_{r=1}^{n_\gamma}\sum_{s=1}^{n_\delta}\eta^\alpha_{j,p}\eta^\beta_{k,q}\eta^\gamma_{l,r}\eta^\delta_{m,s}\nonumber \\ &\times\mathrm{Im}\left[\langle \mathrm{e}^{\mathrm{i}\hat{\mathcal{H}}(t+\tau)}P^\alpha_{j,p} \mathrm{e}^{-\mathrm{i}\hat{\mathcal{H}}t}P^\beta_{k,q} e^{-\mathrm{i}\hat{\mathcal{H}}\tau} P^\gamma_{l,r}P^\delta_{m,s}\rangle\right.\nonumber \\
&\qquad\left. +\langle P^\delta_{m,s}P^\gamma_{l,r} \mathrm{e}^{\mathrm{i}\hat{\mathcal{H}}(t+\tau)}P^\alpha_{j,p} \mathrm{e}^{-\mathrm{i}\hat{\mathcal{H}}t}P^\beta_{k,q} \mathrm{e}^{-\mathrm{i}\hat{\mathcal{H}}\tau} \rangle \right.\nonumber \\
&\qquad\left. -\langle P^\gamma_{l,r} \mathrm{e}^{\mathrm{i}\hat{\mathcal{H}}(t+\tau)}P^\alpha_{j,p} \mathrm{e}^{-\mathrm{i}\hat{\mathcal{H}}t}P^\beta_{k,q} \mathrm{e}^{-\mathrm{i}\hat{\mathcal{H}}\tau}P^\delta_{m,s} \rangle\right.\nonumber \\
&\qquad\left. -\langle P^\delta_{l,r} \mathrm{e}^{\mathrm{i}\hat{\mathcal{H}}(t+\tau)}P^\alpha_{j,p} \mathrm{e}^{-\mathrm{i}\hat{\mathcal{H}}t}P^\beta_{k,q} \mathrm{e}^{-\mathrm{i}\hat{\mathcal{H}}\tau}P^\gamma_{m,s} \rangle\right]\,.
\end{align}

\begin{figure}[t!]

\begin{tikzpicture}
        \node[anchor=south west, inner sep=0] (image) at (0,0){};
        \node[above left] at (-18.5, 0) {(a)};
\end{tikzpicture}

\vspace{-5pt}
\centerline{
\Qcircuit @C=0.75em @R=1.5em{ \lstick{|0\rangle} & \gate{H} & \gate{S} & \ctrl{1} &\ctrl{1}  & \gate{X} & \ctrl{1} &\ctrl{1} & \gate{X}  &\ctrl{1} & \gate{X} & \ctrl{1}  & \gate{H} &\meter\qw\\ \lstick{|\varphi_0\rangle} & \gate{U_\mathrm{G}} & \qw & \gate{P_5} &\gate{P_4} & \qw &\gate{P_0} &\gate{P_1} & \gate{\mathrm{e}^{-\mathrm{i}\h \tau}} &\gate{P_3} & \gate{\mathrm{e}^{-\mathrm{i}\h t}} & \gate{P_2} & \qw & \qw }}

\begin{tikzpicture}
        \node[anchor=south west, inner sep=0] (image) at (0,0){};
        \node[above left] at (-18.5, 0) {(b)};
\end{tikzpicture}

\vspace{-5pt}
\centerline{
\Qcircuit @C=0.75em @R=1.5em{ \lstick{|0\rangle} & \gate{H} & \gate{S} & \ctrl{1} &\ctrl{1}  & \gate{X} & \ctrl{1} &\ctrl{1} & \gate{X}  &\ctrl{1} & \gate{X} & \ctrl{1}  & \gate{H} &\meter\qw\\ \lstick{|\varphi_0\rangle} & \gate{U_\mathrm{G}[\bth^1]} & \qw & \gate{P_5} &\gate{P_4} & \qw &\gate{P_0} &\gate{P_1} & \gate{U_\tau(\bth^3)} &\gate{P_3} & \gate{U_t(\bth^2)} & \gate{P_2} & \qw & \qw }}

\caption{Quantum circuit for measuring the third-order susceptibility eq~\ref{eq:chi3_trans}. (a) Circuit utilizing exact time-evolution gates. The upper horizontal line represents the ancillary qubit, initially in the state $\ket{0}$. $X$, $H$, and $S$ correspond to the Pauli-$X$, Hadamard rotation, and $S$-gate operations on the ancilla. The lower horizontal line denotes the qubit register representing the quantum spin system, initially in a reference product state $\ket{\varphi_0}$, which is rotated to the ground state $\ket{G}=U_\mathrm{G}\ket{\varphi_0}$ by a unitary circuit $U_\mathrm{G}$. Multiple controlled Pauli gates are used to measure different terms in eq~\ref{eq:chi3_trans}. The expectation value $\mathrm{Im}[\bra{G} P_0 P_1 \mathrm{e}^{\mathrm{i}\hat{\mathcal{H}}(t+\tau)} P_2\, \mathrm{e}^{-\mathrm{i}\hat{\mathcal{H}}t} P_3\, \mathrm{e}^{-\mathrm{i}\hat{\mathcal{H}}\tau} P_4 P_5\ket{G}]$ is obtained by Pauli-$Z$ measurement on the ancilla qubit.
(b) The CUL circuit to measure the third-order susceptibility, which is an adaptive variational circuit equivalent to (a). The ground state is prepared using a parameterized circuit $U_\mathrm{G}[\bth^1]$, the state propagation by $\mathrm{e}^{-\mathrm{i}\h \tau}$ is achieved by evolving the angles in $U_\tau [\bth^3]$ and $U_\mathrm{G}[\bth^1]$, and state propagation by $\mathrm{e}^{-\mathrm{i}\h t}$ is achieved by evolving angles in $U_t [\bth^2]$, $U_\tau [\bth^3]$, and $U_\mathrm{G}[\bth^1]$. The parameterized unitaries $U_\mathrm{G}[\bth^1]$, $U_t [\bth^2]$, and $U_\tau [\bth^3]$ are automatically generated using adaptive variational algorithms. 
}
\label{fig4}    
\end{figure}
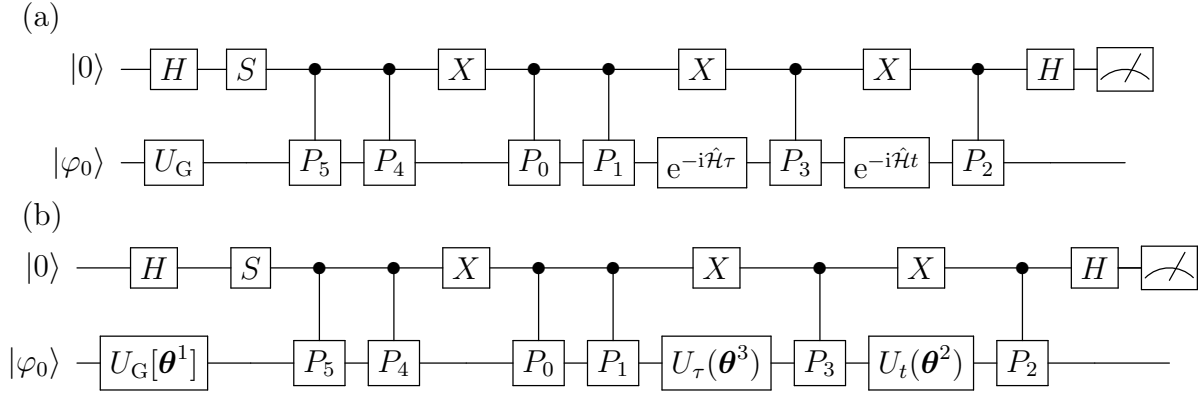

All the terms in eq~\ref{eq:chi3_trans} can be measured using the circuit in Figure~\figref{fig4}{a}. Nevertheless, to reduce the circuit depth for near-term applications, we adopt the generalized CUL circuit shown in Figure~\figref{fig4}{b} for nonlinear correlation function simulations. In the CUL circuit, the state propagation for the system plus ancilla due to $\mathrm{e}^{-\mathrm{i}\h \tau}$ maps to the evolution of parameters in $U_\textrm{G}[\bth^1]$ and $U_\tau[\bth^3]$; while the state propagation due to $\mathrm{e}^{-\mathrm{i}\h t}$ maps to parameter updating in $U_\textrm{G}[\bth^1]$, $U_\tau[\bth^3]$, and $U_t[\bth^2]$. We adopt AVQITE to generate $U_\textrm{G}[\bth^1]$ for ground state preparation, and AVQDS to generate $U_\tau[\bth^3]$ and $U_t[\bth^2]$ for state evolution. More detailed discussions are given in Appendix~\ref{sec:chi_qc} on measuring the different terms within the square brackets of eq~\ref{eq:chi3_trans} using the CUL circuit in Figure~\figref{fig4}{b}. 
Specifically, the CUL circuit measures the imaginary part of
$\bra{G[\bth^1]} P_0 P_1 U_\tau^\dagger[\bth^3] U_t^\dagger[\bth^2] P_2\, U_t[\bth^2] P_3\, U_\tau[\bth^3] P_4 P_5\ket{G[\bth^1]}$.
By setting $P_0=P_1 =I^{\otimes N_q}$, $P_2=P^\alpha_{j,p}$, $P_3=P^\beta_{k,q}$, $P_4=P^\gamma_{l,r}$, and $P_5=P^\delta_{m,s}$ ($P_0=P^\delta_{m,s}$, $P_1=P^\gamma_{l,r}$, $P_2=P^\alpha_{j,p}$, $P_3=P^\beta_{k,q}$, and $P_4=P_5=I^{\otimes N_q}$), the first (second) term within the square brackets of eq~\ref{eq:chi3_trans} is calculated, while the third (fourth) contributions follow by setting $P_0=P_5 =I^{\otimes N_q}$, $P_1=P^{\gamma (\delta)}_{l,r}$, $P_2=P^\alpha_{j,p}$, $P_3=P^\beta_{k,q}$, and $P_4=P^{\delta (\gamma)}_{m,s}$. We note that the CUL circuit in Figure~\figref{fig4}{b} can be simplified for specific settings of $\{P_i\}_{i=0}^5$ gates. 
For instance, if Pauli gate $P_i=I^{\otimes N_q}$, it can be removed trivially. If $P_4 = P_0$, the two controlled $P_4$ and $P_0$ gates can be merged to a single $P_0$ gate.

The calculation of the different terms of eq~\ref{eq:chi3_trans} involves four main steps: (i) Preparation of the ground state of the quantum spin Hamiltonian, $\ket{G[\bth^1]}=U_\mathrm{G}[\bth^1]\ket{\varphi_0}$ where $\ket{\varphi_0}$ is a reference product state, using adaptive variational algorithms like qubit ADAPT-VQE~\cite{MayhallQubitAVQE} or AVQITE~\cite{AVQITE}. This step is analogous to step (i) for the calculation of the Green's function using the quantum circuit in Figure~\figref{fig1}{b}. (ii) Generating and evolving the parameterized unitary circuit $U_\tau(\bth^3(\tau))$ by applying the AVQDS approach discussed in section~\ref{sec:avqds} to propagate the quantum state
\begin{align}
    \ket{\Psi}=\frac{1}{\sqrt{2}}&\left[\ket{1}\otimes P_1 P_0\ket{G[\bth^1(\tau)]}\right. \nonumber \\
    &\left.+\mathrm{i}\ket{0}\otimes P_4 P_5\ket{G[\bth^1(\tau)]}\right]\,,
\label{eq:qc_chi_1}    
\end{align}
where the parameters $\bth^1(\tau)$ also evolve from the ground state solution to leverage their degrees of freedom for dynamics simulations. Here, eq~\ref{eq:qc_chi_1} corresponds to the state in the circuit in Figure~\figref{fig4}{b} after applying the controlled $P_1$ operation. 
(iii) Generating and evolving the parameterized unitary circuit $U_t(\bth^2(t))$ by applying the AVQDS approach to propagate the state
\begin{align}
    \ket{\Psi}=\frac{1}{\sqrt{2}}&\left[\ket{0}\otimes U_\tau(\bth^3(t))P_1 P_0\ket{G[\bth^1(t)]}\right. \nonumber \\
    &\left.+\mathrm{i}\ket{1}\otimes P_3 U_\tau(\bth^3(t))P_4 P_5\ket{G[\bth^1(t)]}\right]\,,
\label{eq:qc_chi_2}     
\end{align}
where $\bth^1(t)$ and $\bth^3(t)$ also vary to facilitate the state propagation using compact circuits.
(iv) Repeat the quantum circuit evaluations in Figure~\figref{fig4}{b} for the different contributions in eq~\ref{eq:chi3_trans}, which in total involves $4 N^4 n_\alpha n_\beta n_\gamma n_\delta$ terms.

\section{Applications \label{sec:apps}}

\subsection{\label{sec:GF_res} Single-particle Green's function of Fermi-Hubbard chains}

\subsubsection*{Model}

To benchmark the performance of the AVQDS approach in calculating the Green’s function as outlined in section~\ref{sec:GF} and to compare the results with alternative approaches in refs~\citenum{endo2020calculation,Gomes2023,Libbi2022},  we study the one-dimensional Fermi-Hubbard model of $N$ sites in its particle-hole-symmetric form:
\begin{align}
\label{eq:Ham}
    \h&=-t\sum_{\langle i,j\rangle,\sigma}\left(\cc_{i,\sigma}\ca_{j,\sigma}+\mathrm{h.c.}\right)\nonumber \\
    &+U\sum_j n_{j,\uparrow}n_{j,\downarrow}-\frac{U}{2}\sum_{j,\sigma}n_{j,\sigma}\,.
\end{align}
Here, $\ca_{j,\sigma}$ denotes the annihilation operator for a fermion with spin 
$\sigma$ at site $j$, $n_{j,\sigma}$ represents the number operator, $t$ denotes the hopping amplitude between neighboring sites, and $U$ corresponds to the on-site interaction strength. We focus on $N=4$ and $N=6$-site chains with open boundary conditions. Specifically, we set the hopping parameter $t=1$ and the on-site interaction is fixed at $U=4.0$. To map Hamiltonian eq~\ref{eq:Ham} to multi-qubit operators, we utilize the Jordan-Wigner transformation~\cite{map_jw}. The encoding of Hamiltonian eq~\ref{eq:Ham} requires $N_q=8$ and $N_q=12$ qubits for $N=4$ and $N=6$-site systems, respectively.

\begin{figure}
\begin{center}
		\includegraphics[scale=0.40]{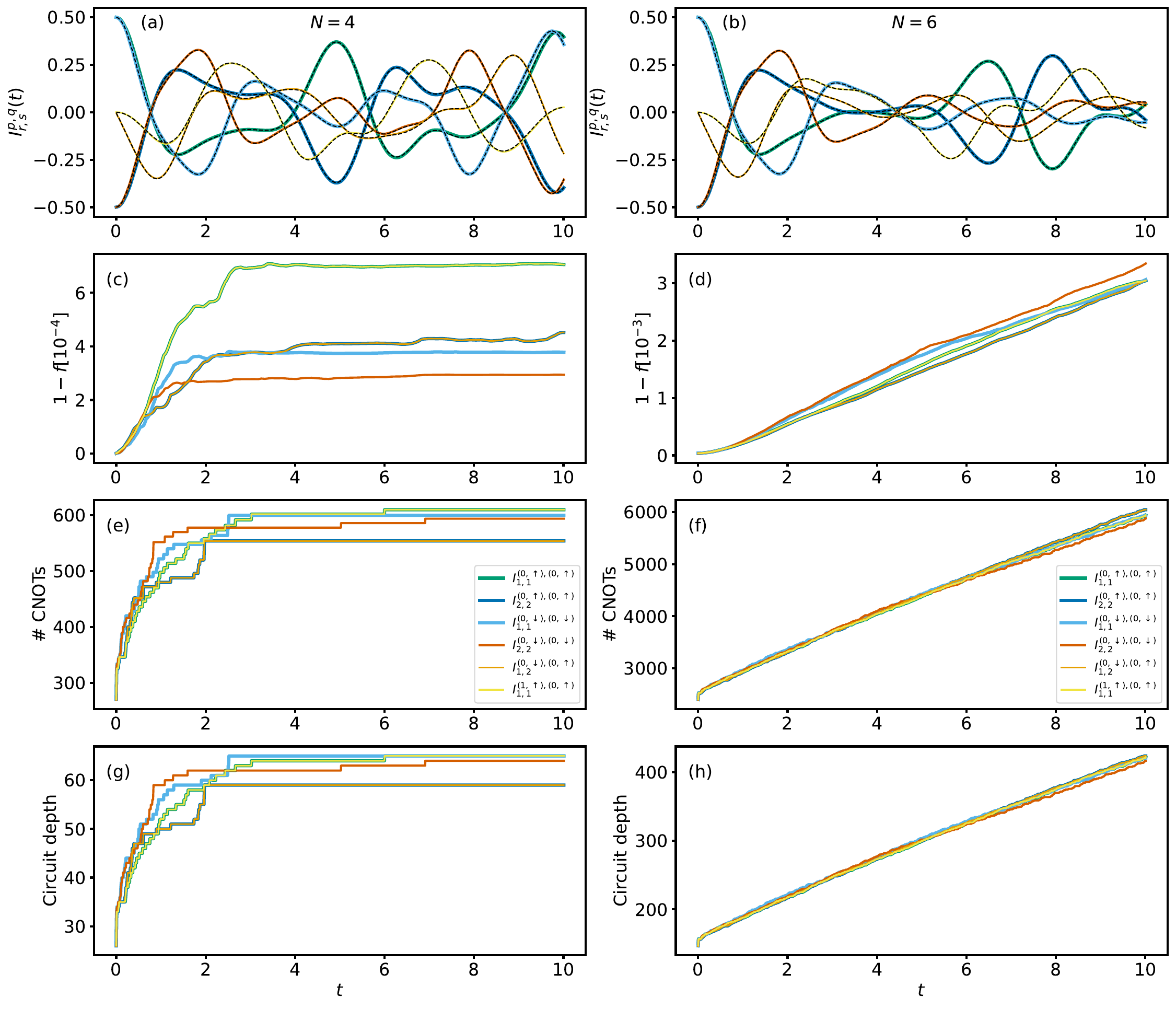}
		\caption{Numerical simulation of the AVQDS approach for computing the single-particle Green's function of Fermi-Hubbard model. Examples of $I^{p,q}_{\alpha,\beta}(t)$ dynamics for six different combinations of $p$, $q$, $\alpha$, and $\beta$, obtained by evaluating the quantum circuit in Figure~\figref{fig1}{b} for Fermi-Hubbard chains with (a) $N=4$- and (b) $N=6$-sites. The results obtained with the AVQDS approach (solid lines) are compared with those of the exact simulations (black dashed lines) obtained via exact diagonalization eq~\ref{eq:ED}. The corresponding infidelities $1-f$ in (c) and (d) demonstrate the high accuracy of AVQDS in calculating the Green's function components $I^{p,q}_{\alpha,\beta}(t)$, achieving a fidelity of at least 99.93~$\%$ for $N=4$ and 99.64~$\%$ for $N=6$. The corresponding number of CNOT gates in (e) and (f) increases from an initial count of 270 (2402) to a maximum of 610 (6148) at the final simulation time of $t=10$ for $N=4$ ($N=6$). The circuit depth in (g) and (h) grows from 26 (147) to a maximum circuit depth of 65 (424) at $t=10$ for $N=4$ ($N=6$). Note that the pair $I^{(0,\uparrow), (0,\uparrow)}_{1,1}$ (green line) and $I^{(1,\uparrow), (0,\uparrow)}_{1,1}$ (yellow line) as well as the pair $I^{(0,\uparrow), (0,\uparrow)}_{2,2}$ (dark blue line) and $I^{(0,\downarrow), (0,\uparrow)}_{1,2}$ (orange line) have the same $P_\beta$ but different $P_\alpha$ in the circuit of Figure~\figref{fig1}{b}. As a result, the circuits for measuring these pairs involve exactly the same parameterized unitaries $U_\mathrm{G}[\bth^1]$ and $U_t[\bth^2]$ in Figure~\figref{fig1}{b}, and therefore the same evolution of the number of CNOTs, circuit depth, and infidelity. However, the dynamics of these pairs in (a) and (b) are distinct due to different $P_\alpha$.}
		\label{fig2} 
\end{center}
\end{figure}

\subsubsection*{Ground state preparation}

The ground state is prepared using the AVQITE approach~\cite{AVQITE}, where the distance between the actual imaginary time-evolved state and the variational ansatz state is minimized in accordance with McLachlan's variational principle~\cite{mclachlan64variational}. Analogous to AVQDS, the McLachlan distance is kept below a predefined threshold during the imaginary time evolution by adaptively appending new parameterized unitaries to the variational ansatz. The generators of the unitaries are chosen from a predetermined operator pool. The obtained ground state ansatz also takes the pseudo-Trotter form eq~\ref{eq:ansatz}, allowing straightforward combination with the AVQDS approach.

In our simulations, we adopt an operator pool derived from the unitary coupled-cluster singles and doubles excitation operators~\cite{MayhallQubitAVQE, smqite}, or the related qubit excitation operators~\cite{yordanov2021qubit}. Specifically, the pool can be written as:
\begin{align}
    \mathcal{P} = \{ \hat{\sigma}_i^p\hat{\sigma}_j^q\} \cup \{ \hat{\sigma}_i^p\hat{\sigma}_j^q \hat{\sigma}_k^r \hat{\sigma}_l^s\}\,,
    \label{eq:sd_pool}
\end{align}
subject to the constraint that only Pauli strings with odd number of $\hat{\sigma}^y$ are included. Here $p, q, r,s \in \{x,y\}$, $i,j,k,l$ run over all qubits, and $\hat{\sigma}^x$ and $\hat{\sigma}^y$ represent Pauli operators. The pool comprises $N_\mathrm{p} = 616$ generators for the model with $N=4$ and $N_\mathrm{p} = 4092$ generators for $N=6$. 
Regarding the reference state $\ket{\varphi_0}$, we opt for a product state with spin-up electrons occupying fermionic orbital sites $1, \cdots, N/2$ and spin-down electrons occupying sites $N/2+1, \cdots, N$, which is subsequently converted to a product state in the qubit representation. Although there are many other forms of reference states~\cite{mukherjee2023comparative}, such as a product state from Hartree-Fock or qubit mean-field calculations~\cite{qcc_scott2018}, this choice is consistent with ref~\citenum{Gomes2023}, which allows a fair comparison on quantum resources between the CUR~\cite{Gomes2023} and our CUL approaches.

\subsubsection*{Simulation results}

We first analyze the performance and quantum resource requirements for the algorithm outlined in section~\ref{sec:GF}. For the operator pool in the AVQDS calculations, we utilize the Hamiltonian operator pool, which encompasses all individual Pauli strings contained in the qubit representation of the Hamiltonian. As a result, the pool comprises $N_\mathrm{p}=16$ ($N_\mathrm{p}=26$) operators for $N=4$ ($N=6$) model. To compute the Green's function $G^\mathrm{R}_{p,q}(t)$ using eq~\ref{eq:Gr_trafo}, the quantum circuit in Figure~\figref{fig1}{b} needs to be calculated for all $\alpha,\beta$-components in eq~\ref{eq:Gr_trafo}. Since the expression of the creation or annihilation operators according to eq~\ref{eq:JW} consists of $n=2$ terms after the Jordan-Wigner transformation, computing the Green's function via eq~\ref{eq:Gr_trafo} involves $n^2=4$ simulations of the quantum circuit in Figure~\figref{fig1}{b} for a given $p,q$-component. 

Figure~\figref{fig2}{a} and \figref{fig2}{b} show examples of the  $I^{p,q}_{\alpha,\beta}$ dynamics for different combinations of $p$, $q$, $\alpha$, and $\beta$, obtained by evaluating the quantum circuit presented in Figure~\figref{fig1}{b} for the $N=4$- and $N=6$-site Fermi-Hubbard chains. The results of the AVQDS approach (solid lines) are plotted together with the corresponding outcomes of the exact simulation (black dashed lines) obtained via exact diagonalization eq~\ref{eq:ED}. The AVQDS results closely align with those of the exact simulations for all presented $I^{p,q}_{\alpha,\beta}$. To quantify the precision of the AVQDS calculations, we plot the corresponding infidelity in Figure~\figref{fig2}{c} and \figref{fig2}{d}, defined as $1-f=1-|\langle \Psi[\bth(t)]|\Psi_\mathrm{exact}(t)\rangle|^2$. Here, $|\Psi_\mathrm{exact}[t]\rangle = I\otimes \mathrm{e}^{-\mathrm{i}\h t}\ket{\Psi}$ with $\ket{\Psi}$ defined in eq~\ref{eq:qc1} is calculated using exact diagonalization eq~\ref{eq:ED}, while $|\Psi[\bth(t)]\rangle$ represents the corresponding variational wavefunction obtained through AVQDS. The infidelity remains below $7.1\times 10^{-4}$ ($3.6\times 10^{-3})$ within the studied time window for the $N=4$-site ($N=6$-site) simulations. This demonstrates that on a statevector simulator AVQDS can accurately calculate the real-time Green's function components $I^{p,q}_{\alpha,\beta}$ using the quantum circuit presented in Figure~\figref{fig1}{b}.

The above results are obtained from noiseless statevector simulations and therefore neglect the effects of gate errors as well as finite sampling of the quantum state when measuring the matrix $M$ and vector $V$ in eqs~\ref{eq:M} and \ref{eq:V}. Thus, it is important to estimate the quantum resources necessary for AVQDS calculations on noisy intermediate-scale quantum devices. Our focus here is on the quantum circuit complexity quantified by the number of two-qubit CNOT gates. To simplify the analysis, we assume full qubit connectivity, as found in trapped-ion architecture, where implementing the multi-qubit rotation gate $\mathrm{e}^{-\mathrm{i}\theta\hat{\A}}$ demands $2(p-1)$ CNOT gates for a Pauli string $\hat{\A}$ containing $p$ Pauli operators. Figure~\figref{fig2}{e} and \figref{fig2}{f} show the growth of the CNOT number with time for the $I^{p,q}_{\alpha,\beta}$ simulations presented in Figure~\figref{fig2}{a} and \figref{fig2}{b}. The AVQITE ground-state preparation achieves an infidelity of approximately $10^{-6}$ and $3.6\times 10^{-5}$ for $N=4$ and $N=6$, respectively. The number of variational parameters $N_\theta$ required for the ground state preparation is 65 for $N=4$ and and 551 for $N=6$. Specifically, the associated variational ansatz eq~\ref{eq:ansatz} incorporates 30 two- and 35 four-qubit rotation gates for $N=4$, resulting in 270 CNOTs, and 226 two- and 325 four-qubit rotation gates for $N=6$, yielding 2402 CNOTs. The number of CNOTs increases rapidly up to $t\approx 3$ for the $N=4$ simulations, before saturating at prolonged duration, while it exhibits a sublinear increase during the entire simulation up to $t=10$ for the $N=6$ simulations. The number of CNOTs rises from the initial 270 (2402) for ground state preparation to a maximum number of 610 (6148) at the final simulation time of $t=10$ for $N=4$ ($N=6$). The resulting number of CNOTs at the final simulation time is comparable for the different components $I^{p,q}_{\alpha,\beta}$. In addition to the CNOT count and number of qubits, the circuit depth determined by the number of layers represents a critical metric for noisy intermediate-scale quantum devices. Figure~\figref{fig2}{g} and \figref{fig2}{h} present the evolution of the circuit depth with time for the $I^{p,q}_{\alpha,\beta}$ dynamics presented in Figure~\figref{fig2}{a} and \figref{fig2}{b}. The depths of the state-preparation circuits are 26 and 147 for $N=4$ and $N=6$, respectively. During the time evolution, the circuit depth grows to a maximum of 65 (424) at the final simulation time of $t=10$ for $N=4$ ($N=6$). The observed saturation of the circuit depth at long simulation times for $N=4$ and sublinear increase for $N=6$ point to a significant advantage compared to Trotterization, where the circuit depth is determined by the number and weight of the Hamiltonian terms and grows linearly with the number of time steps~\cite{AVQDS,mootz2023twodimensional}. We also note that on today's quantum processors, circuits of up to 60 layers of two-qubit gates can be successfully executed, as demonstrated in ref\citenum{kimEvidenceUtilityQuantum2023}, which is close to the circuit depth required for simulating the Green's function of the four-site Hubbard model in Figure~\ref{fig2}.

For comparison, if variational quantum dynamics simulation with HVA is used in the CUL Green's function calculations, the number of two-qubit rotation gates required for a single layer HVA is $8N^{3/2}+N-4\sqrt{N}$ for the Fermi-Hubbard model with $N$ sites and open boundary conditions~\cite{endo2020calculation}. For instance, with $N=4$ and 16 layers of HVA as used in ref~\citenum{endo2020calculation}, the estimated number of two-qubit rotation gates is 960. Consequently, variational quantum dynamics simulation with a 16-layer HVA exceeds the maximum number of required CNOT gates of our AVQDS approach by approximately 50~$\%$, despite yielding less accurate results at long simulation times compared to our results based on the AVQDS approach. This demonstrates that the calculations of the Green's function with AVQDS lead to much shallower quantum circuits relative to HVA. To compare with the CUR approach~\cite{Gomes2023}, we subtract the CNOTs in the ground state preparation circuit from the CNOT counts in Figure~\figref{fig2}{e}, as only the number of CNOTs accumulated during the time evolution is presented in ref~\citenum{Gomes2023}. After this adjustment, the maximum increase in the CNOT count during the time evolution is 340 within the studied time window of $[0, 10]$, which is about the same as that in ref~\citenum{Gomes2023}. We note that for the overlap test in CUR method, an additional $2N_\textrm{p}$ CNOTs are needed, which is about $2\times 70=140$. Here $N_\textrm{p}$ is the number of parameters introduced in the single-state AVQDS calculations in the CUR approach. Comparable CNOT counts but deeper quantum circuits are reported for CUR calculations using HVA for variational state propagation~\cite{Libbi2022}.

\begin{figure}[t!]
\begin{center}
		\includegraphics[scale=0.43]{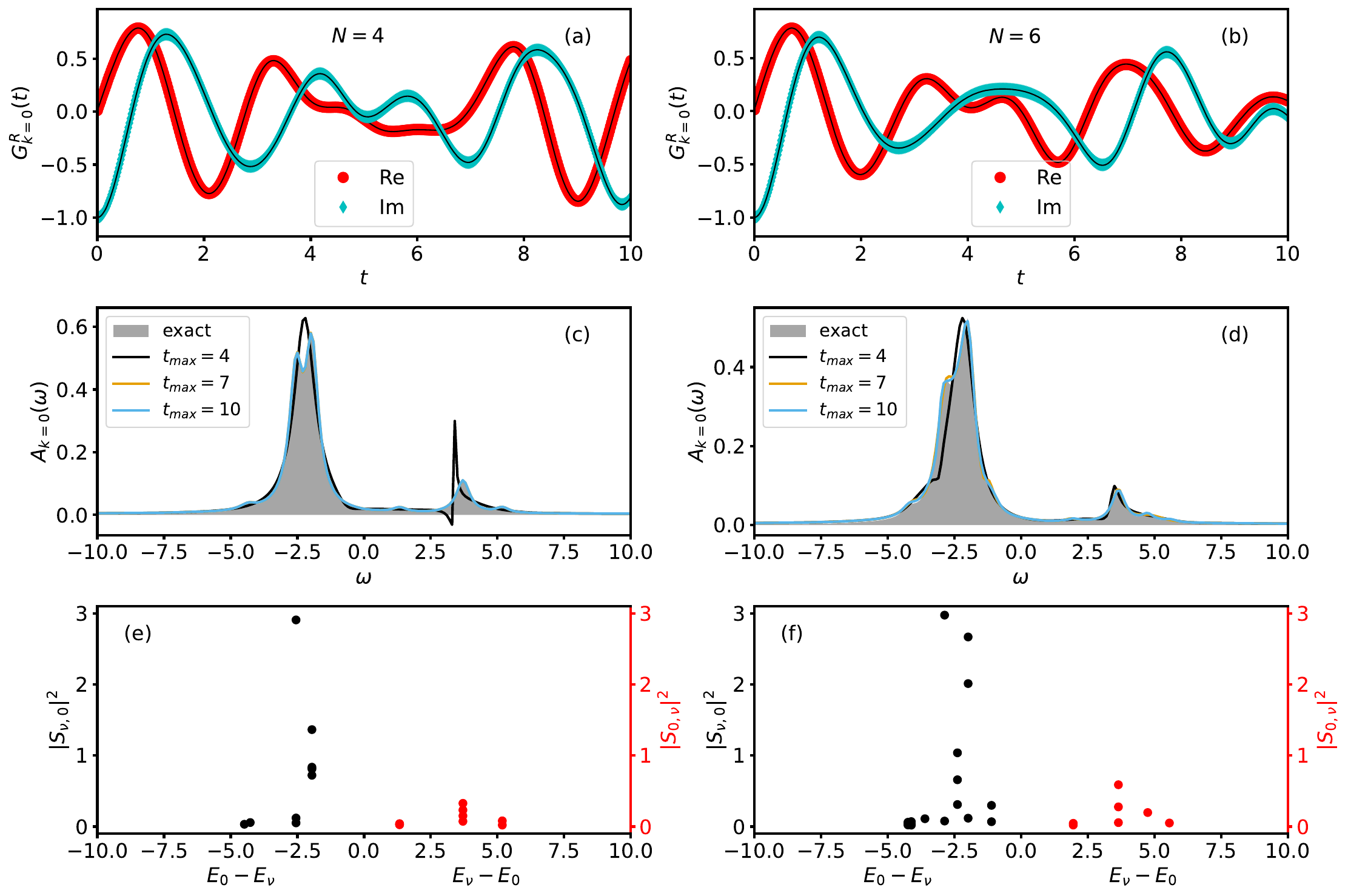}
		\caption{Single-particle Green's function in momentum space and spectral function. Dynamics of the real and imaginary parts of $G^\mathrm{R}_{k}(t)$ at momentum $k=0$ for Fermi-Hubbard model with (a) $N=4$- and (b) $N=6$-sites. The real part (red circles) and imaginary part (cyan diamonds) of $G^\mathrm{R}_{k=0}(t)$ obtained with the AVQDS approach agree well with the corresponding results of the exact simulations (solid black lines).  (c), (d) Spectral function $A_{k=0}(\omega)$ obtained by Fourier transforming the dynamics presented in (a) and (b) using the Pad\'e approximation and calculating eq~\ref{eq:Ak}. The result is shown for three different $t_\mathrm{max}$ used within the Pad\'e approximation. As a comparison, the exact result for the spectral function based on eq~\ref{eq:GL} is plotted as a shaded area. $t_\mathrm{max}=7$ is sufficient to accurately reproduce the main features in $A_{k=0}(\omega)$. (e), (f) $|S_{\nu,0}|^2$ (black dots) and $|S_{0,\nu}|^2$ (red dots) as a function of the energy differences $E_0-E_\nu$ and $E_\nu - E_0$, respectively, for (e) $N=4$ and (f) $N=6$. Only the dominant transition amplitudes with $|S_{\nu,\mu}|^2 > 0.02$ are shown. The peaks in the spectral functions originate from transitions between the ground state with energy $E_0$ to the excited states with energies $E_\nu$, and vice versa.}
		\label{fig3} 
\end{center}
\end{figure}

It is important to note that the quantum resource estimates for Green's function calculations using the proposed quantum computing approach include the cost of ground state preparation as well as that of state propagation for the combined ancilla and physical qubit system. With a focus on near-term quantum algorithms, we utilize adaptive variational quantum algorithms for the calculations leveraging the problem-specific compact ans\"atze they automatically generate. 
Although the  number of parameters needed for the variational wavefunction for large-size systems can be large, we emphasize that the adaptive variational algorithms we adopt only evolve the parameters along the real or imaginary time axis, without requiring explicit high-dimensional parameter optimizations. The adaptive algorithms construct and adjust compact ans\"atze along the dynamical path. This ensures that the exact state evolution can be followed closely, which often allows substantially more time steps to be taken than for a fixed ansatz with a restricted number of variational parameters. 
 
 Due to the nature of the adaptive algorithms, it is hard to derive a generic analytical upper bound for the required quantum resources and the practical performance may be problem-dependent. Nevertheless, substantial efforts have been devoted to numerically benchmarking the adaptive variational algorithms. For instance, highly compact ground state ans\"atze with much lower circuit depth than the problem-agnostic unitary coupled cluster ansatz have been demonstrated with AVQITE applications to molecular systems~\cite{AVQITE}. AVQDS has been shown capable of simulating quantum dynamics with circuits of depth over two orders of magnitude less than standard Trotter circuits~\cite{AVQDS,mootz2023twodimensional}. More importantly, the system-size scaling of the quantum circuit complexity characterized by the number of CNOT gates has also been numerically studied with a set of spin models, including high-spin systems. Although a definitive asymptotic scaling cannot be extracted due to limited system sizes, the numerical benchmarks suggest a polynomial growth of quantum circuit depth for larger-size systems, in contrast to the exponential scaling of exact diagonalization methods. The adaptive variational quantum algorithms can naturally leverage the computing power of quantum processing units, namely the efficient representation of the quantum many-body state and unitary state evolution. This renders the proposed quantum computing approach advantageous over classical computing for simulating quantum many-body systems.

\subsubsection*{Momentum-space Green's function and spectral function}

To further demonstrate the accuracy of CUL Green's function calculations using AVQDS, we study the retarded Green's function in momentum space, which has the following form:
\begin{align}
\label{eq:GRk2}
    G^\mathrm{R}_{k,\sigma}&=\frac{1}{N}\sum_{i, j=1}^{N}G^\mathrm{R}_{(i,\sigma),(j,\sigma)}\mathrm{e}^{-\mathrm{i} k(i-j)}\,.
\end{align}
To calculate eq~\ref{eq:GRk2} using eq~\ref{eq:Gr_trafo}, $I^{(i,\sigma),(j,\sigma)}_{\alpha,\beta}(t)$ is evaluated for all $i$, $j$, $\sigma$, $\alpha$, and $\beta$. In general, the $t$-stepping in the dynamics of $I^{(i,\sigma),(j,\sigma)}_{\alpha,\beta}(t)$ is not equidistant due to the dynamically adjusted $\delta t$ in the AVQDS approach and is also component-dependent. In practice, we obtain $I^{(i,\sigma),(j,\sigma)}_{\alpha,\beta}(t)$ on a uniform $t$-mesh using linear interpolation before calculating $G^\mathrm{R}_{k,\sigma}$ via eq~\ref{eq:GRk2}. For simplicity, we set $G^\mathrm{R}_{k}\equiv G^\mathrm{R}_{k,\sigma}$ due to spin rotation symmetry. Figure~\figref{fig3}{a} and \figref{fig3}{b} show the real and imaginary parts of $G^\mathrm{R}_{k}(t)$ at momentum $k=0$ for Fermi-Hubbard chains with $N=4$- and $N=6$-sites, respectively. The real and imaginary parts of $G^\mathrm{R}_{k=0}$ obtained with the AVQDS approach are shown as red circles and cyan diamonds, respectively, which agree well with the exact simulation results represented by solid black lines.

We next study the spectral function, which is defined by
\begin{align}
\label{eq:Ak}
    A_k(\omega)=-\frac{1}{\pi}\mathrm{Im}\left[G^\mathrm{R}_k(\omega)\right]\,,
\end{align}
where $G^\mathrm{R}_k(\omega)$ is the Fourier transform of $G^\mathrm{R}_k(t)$, 
\begin{align}
\label{eq:GR_FT}
    G^\mathrm{R}_k(\omega)=\int_{-\infty}^{\infty}\mathrm{d}t\,\mathrm{e}^{\mathrm{i}(\omega+\mathrm{i}\varepsilon)t}G^\mathrm{R}_k(t)\,,
\end{align}
with infinitesimal $\varepsilon$ to guarantee convergence of the integral.
The calculation of eq~\ref{eq:GR_FT} is challenging, as the accurate computation of the spectral function requires long simulation times. Reference~\citenum{Yu2024} presents a quantum computing approach for accurately calculating spectral functions by reducing noise and extending time-domain results through denoising the imaginary time response functions using Hankel projections. Here, however, we adopt the Pad\'e approximation for spectral analysis~\cite{bruner2016accelerated, Gomes2023}, which has been demonstrated to accelerate the convergence of Fourier transforms with simulation time. To proceed, one first casts the discrete form of the Fourier transform eq~\ref{eq:GR_FT} as a power series expansion:
\begin{align}
G^\mathrm{R}_k(\omega)=\sum_{n=0}^{N_\textrm{T}}G^\mathrm{R}_k(t_n)z^n\,, \label{eq:pse}
\end{align}
where $t_n=n \delta t$ is the $n$th point in a uniform time mesh $[0, T]$ with time step size $\delta t = T/N_\textrm{T}$ with $N_\textrm{T}$ even, and $z^n=\left(\mathrm{e}^{\mathrm{i}(\omega+\mathrm{i}\varepsilon)\delta t}\right)^n$. We choose a diagonal Pad\'e approximation~\cite{Pade2016}, where $G^\mathrm{R}_k(\omega)$ is expressed as a ratio of two polynomials of equal order:
\begin{align}
\label{eq:Pade}
    G^\mathrm{R}_k(\omega)=\frac{\sum_{n=0}^{N_\textrm{T}/2}a_n z^n}{1+\sum_{n=1}^{N_\textrm{T}/2}b_n z^n}\,.
\end{align}
The coefficients of these polynomials, $a_n$ and $b_n$, are obtained by solving the system of linear equations obtained from matching orders of $z^n$ in eqs~\ref{eq:pse} and \ref{eq:Pade}, as implemented in SciPy~\cite{2020SciPy-NMeth}. The resulting rational function eq~\ref{eq:Pade} can then be used as an approximation to the original function. Since the coefficients are independent of frequency, $G^\mathrm{R}_k(\omega)$ can be calculated for any frequency based on eq~\ref{eq:Pade}, in contrast to the Fast Fourier transform (FFT) where the spectral resolution is determined by the maximum simulation time $t_\mathrm{max}$. However, the accuracy of the Pad\'e approximant depends on $t_\mathrm{max}$, as illustrated below.

Figure~\figref{fig3}{c} and \figref{fig3}{d} show the spectral function $A_{k=0}(\omega)$ obtained by calculating the Fourier transform of the real-time components presented in Figure~\figref{fig3}{a} and \figref{fig3}{b}, using the Pad\'e approximation with a damping factor of $\varepsilon=0.3$ (tied to $t_\mathrm{max}$) and evaluating eq~\ref{eq:Ak}. The results are shown for three different values of $t_\mathrm{max}$ used in the Pad\'e approximation. For comparison, we also present the exact result for the spectral function (shaded area), obtained using the Lehmann representation of the Green's function, as derived in Appendix~\ref{sec:Lehmann}. The simulated spectral function based on the AVQDS approach agrees well with the exact $A_{k=0}(\omega)$. In particular, a simulation time of $t_\mathrm{max}=7$ is already sufficient to accurately reproduce the main features in $A_{k=0}(\omega)$ for the damping factor $\varepsilon=0.3$ considered in the Fourier transformation eq~\ref{eq:GR_FT}. Compared to the discrete Fourier transform eq~\ref{eq:pse}, the Pad\'e approximation requires a smaller $t_\mathrm{max}$ to accurately calculate the spectral function, as demonstrated in Appendix~\ref{sec:FT}.

To identify the origin of the peaks in the spectral function $A_{k=0}(\omega)$, we study the Lehmann representation of the Green's function in more detail. For $k=0$, the Lehmann representation of the Green's function eq~\ref{eq:GRk} derived in Appendix~\ref{sec:Lehmann} simplifies to
\begin{align}
\label{eq:GL}
    &G^\mathrm{R}_{k=0}(\omega)=\frac{1}{N}\sum_\nu\left[\frac{|S_{0,\nu}|^2}{E_0-E_\nu+\omega+\mathrm{i}\,\varepsilon}\right. \nonumber \\ &\qquad\qquad\qquad\qquad\left.+\frac{|S_{\nu,0}|^2}{E_\nu-E_0+\omega+\mathrm{i}\,\varepsilon}\right]\,,\nonumber \\
    &S_{\nu,\mu}\equiv\sum_p T^p_{\nu,\mu}\,,
\end{align}
where $T^p_{\mu,\nu}\equiv\bra{\Psi_\mu}\ca_p\ket{\Psi_\nu}$ are the transition matrix elements between eigenstates $\ket{\Psi_\nu}$ and $\ket{\Psi_\mu}$ of the Hamiltonian eq~\ref{eq:Ham}. From this expression, it is evident that the peaks in the spectral function arise from transitions between the ground state with energy $E_0$ to the excited states with energies $E_\nu$ by adding an electron [first term in $G^\mathrm{R}_{k=0}$, eq~\ref{eq:GL}] or removing an electron [second term in $G^\mathrm{R}_{k=0}$, eq~\ref{eq:GL}], and vice versa. The spectral weight of these peaks in $A_{k=0}(\omega)$ is determined by $S_{\mu,\nu}$. Figure~\figref{fig3}{e} and \figref{fig3}{f} present $|S_{\nu,0}|^2$ (black dots) and $|S_{0,\nu}|^2$ (red dots) as a function of the energy differences $E_0-E_\nu$ and $E_\nu - E_0$, respectively, for $N=4$ (Figure~\figref{fig3}{e}) and $N=6$ (Figure~\figref{fig3}{f}). In the case of $N=4$, the energetically close peaks around $\omega\approx -2.2$ in the spectral function in Figure~\figref{fig3}{c} result from transitions between degenerate excited states $\nu=20,...,23$ to the ground state as well as from the transition of the degenerate states $\nu=32,...,35$ to the ground state. The states $\nu=20,...,23$ have energy $E_0-E_\nu=-1.955$ while $E_0-E_{\nu=32,...,35}=-2.55$. The dominant signal at positive frequencies, $\omega\approx 3.7$, in Figure~\figref{fig3}{c} results from transitions between the ground state and degenerate excited states $\nu=51,...,54$ with energy $E_\nu-E_0=3.71$. For $N=6$, the peaks around $\omega\approx -2.5$ in the spectral function in Figure~\figref{fig3}{d} stem from transitions between excited states $\nu=49,...,52$, $\nu=72,...,75$, and $\nu=115,...,118$ to the ground state. The states $\nu=49,...,52$ ($\nu=72,...,75$) have energy $E_0-E_\nu=-1.98$ ($E_0-E_\nu= -2.38$) while $E_0-E_{\nu=115,...,118}=-2.87$. The dominant signal at positive frequencies, $\omega\approx 3.6$, in Figure~\figref{fig3}{d} mainly results from transitions between the ground state and degenerate excited states $\nu=223,...,226$ with energy $E_\nu-E_0=3.65$. As a result, the spectral function provides direct insights into the energy levels and excitations of fermions in the studied Fermi-Hubbard model.

To characterize the excited states contributing to the dominant peaks in the spectral functions in more detail, we calculate the expectation values of the total spin operator $\hat{S}^2 = (\hat{S}^x)^2 + (\hat{S}^y)^2 + (\hat{S}^z)^2$ for these states. Here, the spin operators expressed in terms of fermionic creation ($\cc$) and annihilation ($\ca$) operators are given by
\begin{align}
\hat{S}^\alpha = \sum_{i=1}^N \sum_{s,s'=\uparrow,\downarrow}\cc_{i,s} \left( \sigma^\alpha_{s,s'} \right) \ca_{i,s'},
\label{eq:S_a}
\end{align}
where $\alpha = x, y, z$ denotes three spatial components, $\sigma^\alpha$ are the Pauli matrices, and $i$ is a site index. The calculation of $\langle\hat{S}^2\rangle =\bra{\Psi_\nu}\hat{S}^2\ket{\Psi_\nu}=S(S+1)$ then yields the total spin quantum number $S$ of the eigenstate $\ket{\Psi_\nu}$. To determine $S$, we obtain the eigenstates $\ket{\Psi_\nu}$ via exact diagonalization of the Hamiltonian and then calculate $\bra{\Psi_\nu}\hat{S}^2\ket{\Psi_\nu}$ after transforming the fermionic operators in eq~\eqref{eq:S_a} to Pauli operators using the Jordan-Wigner transformation. To further characterize the excited states, we also calculate the particle number of the state $\ket{\Psi_\nu}$, which follows from
\begin{align}
N_e = \sum_{i=1}^{N}\sum_{\sigma=\uparrow,\downarrow} \bra{\Psi_\nu} \cc_{i,\sigma} \ca_{i,\sigma}\ket{\Psi_\nu}\,.
\end{align}

For the $N = 4$ site Fermi-Hubbard chain, we find that four degenerate excited states contribute to each of the three dominant peaks in the spectral function (two dominant peaks at negative frequencies, $\omega\approx -1.96$ and $\omega\approx -2.55$, and one dominant peak at positive frequency, $\omega\approx 3.71$). Two of these four excited states have particle number of $N_e=3$, while the remaining two have $N_e=5$. It is important to note that the peaks in the spectral function result from transitions between the ground state and excited states with one electron removed or added, as the ground state has $N_e = 4$ due to the considered half-filling. For these states, the possible values for $S$ are $S = \frac{1}{2}$ and $S = \frac{3}{2}$ for $N_e=3$ and $S = \frac{1}{2}$, $S = \frac{3}{2}$, and $S = \frac{5}{2}$ for $N_e=5$. The calculated $\langle\hat{S}^2\rangle$ of all excited states contributing to the three dominant signals in the spectral function is $S(S+1) = \frac{3}{4}$. This corresponds to a spin quantum number $S = \frac{1}{2}$ and a degeneracy of $2S + 1 = 2$ for both the $N_e=3$ and 
$N_e=5$ states, giving a total degeneracy of four. 

Similarly, for $N = 6$ sites, four states contribute to each of the four dominant peaks in the spectral function (three dominant peaks at negative frequencies ($\omega\approx -2.87, -2.38, -1.98$) and one dominant signal at positive frequency ($\omega\approx 3.65$)). Two of these four excited states have a particle number of $N_e=5$ while the remaining two have $N_e=7$. These states also have a spin quantum number $S = \frac{1}{2}$ and a degeneracy of $2S + 1 = 2$ for both the $N_e=5$ and $N_e=7$ states, resulting in a total degeneracy of four.

\subsubsection*{Calculation of the spectral function with Prony approximation and compressive sensing}
In the above Green's function analysis using the Pad\'e approximation~\ref{eq:Pade}, finite broadening with $\varepsilon=0.3$ is applied to mimic the infinite lattice with continuous energy bands. However, rigorously speaking, our calculations being carried out for finite systems at zero temperature implies zero broadening for the energy levels. Therefore, it is interesting to investigate and compare approaches in addition to the Pad\'e approximation for modelling delta-function peaks. Specifically, we include the Prony approximation~\cite{Gazizova2024,Prony} and compressive sensing~\cite{com-sens1,com-sens2}. In the Prony approximation, a signal is decomposed into a sum of damped exponentials. When applied to the Green's function $G(t)$, this method allows the identification of the complex frequencies (poles) and corresponding residues that characterize the system's spectral properties. Meanwhile, compressive sensing is a technique used to recover sparse signals from incomplete or noisy data. Detailed discussions of how these methods are employed to obtain the spectral function from the Green's function in the time domain are given in Appendices~\ref{sec:Prony} and~\ref{sec:CS} for the Prony approximation and compressive sensing, respectively.

\begin{figure*}[t!]
\begin{center}
		\includegraphics[scale=0.45]{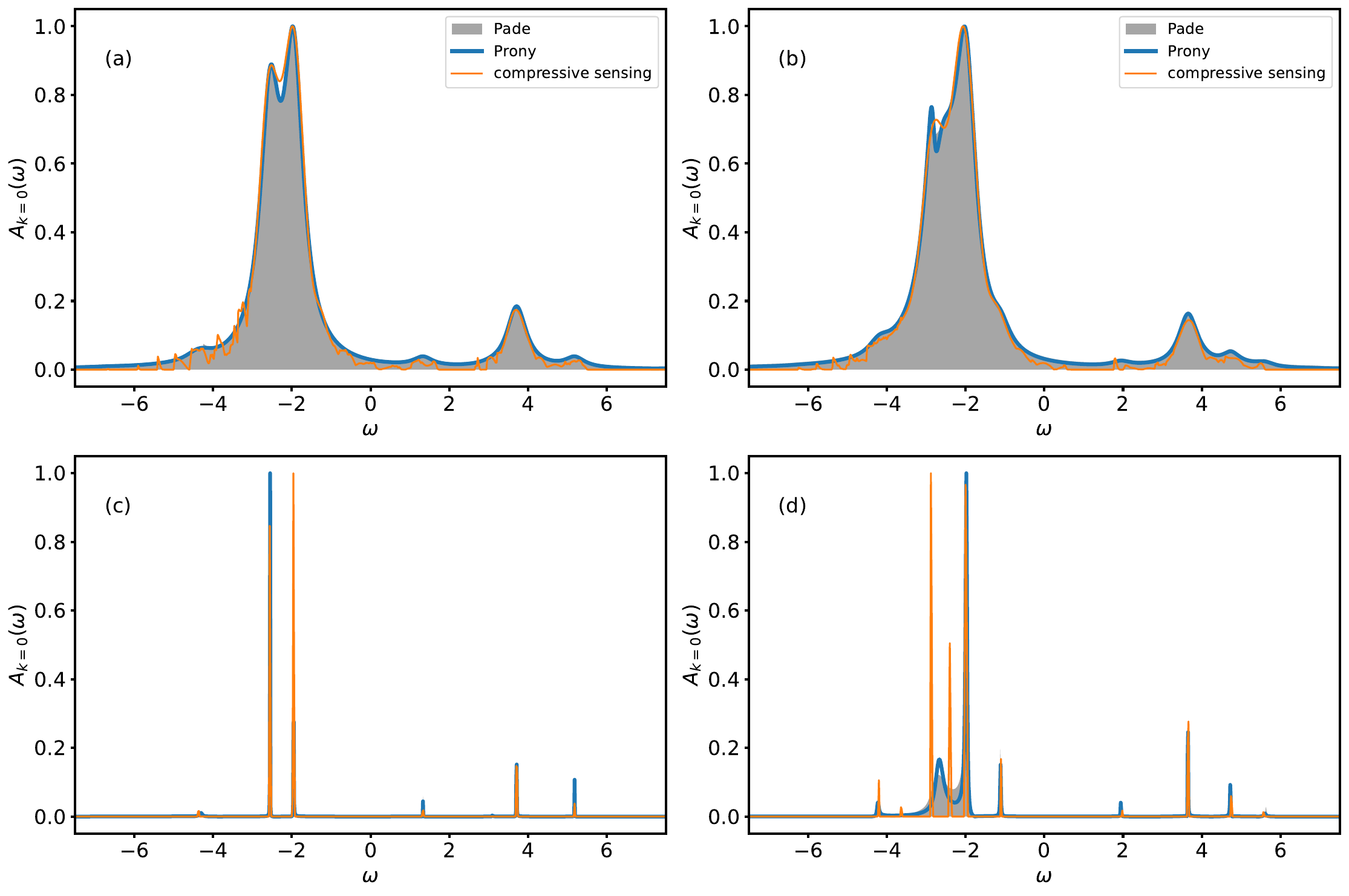}
\caption{Comparison of different signal processing techniques for spectral function calculation. (a), (b) Spectral function $A_{k=0}(\omega)$ obtained by Fourier transforming the dynamics presented in Figure~\figref{fig3}{a} and \figref{fig3}{b}, using a damping factor of $\varepsilon=0.3$ in the Fourier transformation. The result of the Pad\'e approximation (shaded area) is shown together with the results of Prony approximation (blue line) and compressive sensing (orange line). (c), (d) The corresponding results without broadening ($\varepsilon=0$) for (c) $N=4$ and (d) $N=6$. 
}
		\label{fig10} 
\end{center}
\end{figure*}

We compare the performances of the Pad\'e approximation, the Prony approximation, and compressive sensing in calculating the spectral function $A_{k=0}(\omega)$ for the $N=4$ and $N=6$-site Fermi-Hubbard chains. For the Pad\'e and Prony approximations we use uniform time and frequency meshes with $N_t=501$ time steps and $N_\omega=1001$ frequency points. For the compressive sensing method we consider sparser uniform time and frequency meshes of sizes $N_t=101$ and $N_\omega=201$. Figure~\figref{fig10}{a} and \figref{fig10}{b} show the spectral functions $A_{k=0}(\omega)$ obtained by calculating the Fourier transform of the real-time Green's function components presented in Figure~\figref{fig3}{a} and \figref{fig3}{b}, using a damping factor of $\varepsilon=0.3$ in the Fourier transformation. The result of the Pad\'e approximation (shaded area) is compared with the results of the Prony approximation (blue line) and compressive sensing (orange line). The corresponding results without broadening ($\varepsilon=0$) are presented in Figure~\figref{fig10}{c} and \figref{fig10}{d}. 
For finite broadening, all three methods produce similar results, with the spectral functions showing all dominant signals consistent with Figure~\figref{fig3}{e} and \figref{fig3}{f}. Nevertheless, the results from compressive sensing exhibit multiple small artificial spikes. In contrast, for zero broadening (Figure~\figref{fig10}{c} and \figref{fig10}{d}), compressive sensing produces the most accurate results for the delta peaks when compared to Figure~\figref{fig3}{e} and \figref{fig3}{f}, highlighting its effectiveness in resolving sharp spectral features. In particular, in contrast to the Pad\'e and Prony approximations, the compressive sensing method generates $A_{k=0}(\omega)$ with spectral weights more consistent with the exact analysis results shown in Figure~\figref{fig3}{e} and \figref{fig3}{f}. For instance, considering the three main spectral peaks at $\omega = -2.55, -1.96, 3.71$ of N=4-site model, the ratio of spectral weights is 1:0.30:0.03 for the Pad\'e approximation and 1:0.28:0.16 for the Prony approximation; while the ratio becomes 1:1.18:0.17 with compressive sensing, which is much closer to the ratio 1:1.21:0.25 obtained from the exact analysis. This demonstrates that the compressive sensing approach is most effective for resolving delta peaks.

\subsection{Single-particle Green's function of molecule LiH \label{sec:molecule}}

The method introduced in section~\ref{sec:GF} can naturally be used to calculate the single-particle Green's functions of molecules, providing valuable information about molecular ionization potential, electron affinities and excitation energies. To demonstrate this, we compute the Green's function of the molecule LiH. We use the PySCF quantum chemistry package to generate the molecular Hamiltonian~\cite{pyscf_sun2018}. We adopt the minimal STO-3G basis set, treating the $1s$-orbital of Li as a core orbital, with a bond length of 1.547~\AA\ close to equilibrium. The resulting molecular Hamiltonian expressed in the Hartree-Fock molecular orbital basis consisting of 5 spatial orbitals with 2 electron filling, is transformed to the qubit representation using the Jordan-Wigner encoding requiring in total 10 qubits. Similar to calculations of the Fermi-Hubbard model, we use AVQITE to prepare the ground state for LiH. The AVQITE calculation adopts the unitary coupled-cluster singles and doubles excitation operator pool~\ref{eq:sd_pool}, which consists of $N_\mathrm{p} = 144$ generators. As reference state we choose the Hartree-Fock product state. For the operator pool in the AVQDS calculations, we consider the Hamiltonian operator pool, which comprises $N_\mathrm{p}=276$ operators.

\begin{figure*}[t!]
\begin{center}
		\includegraphics[scale=0.55]{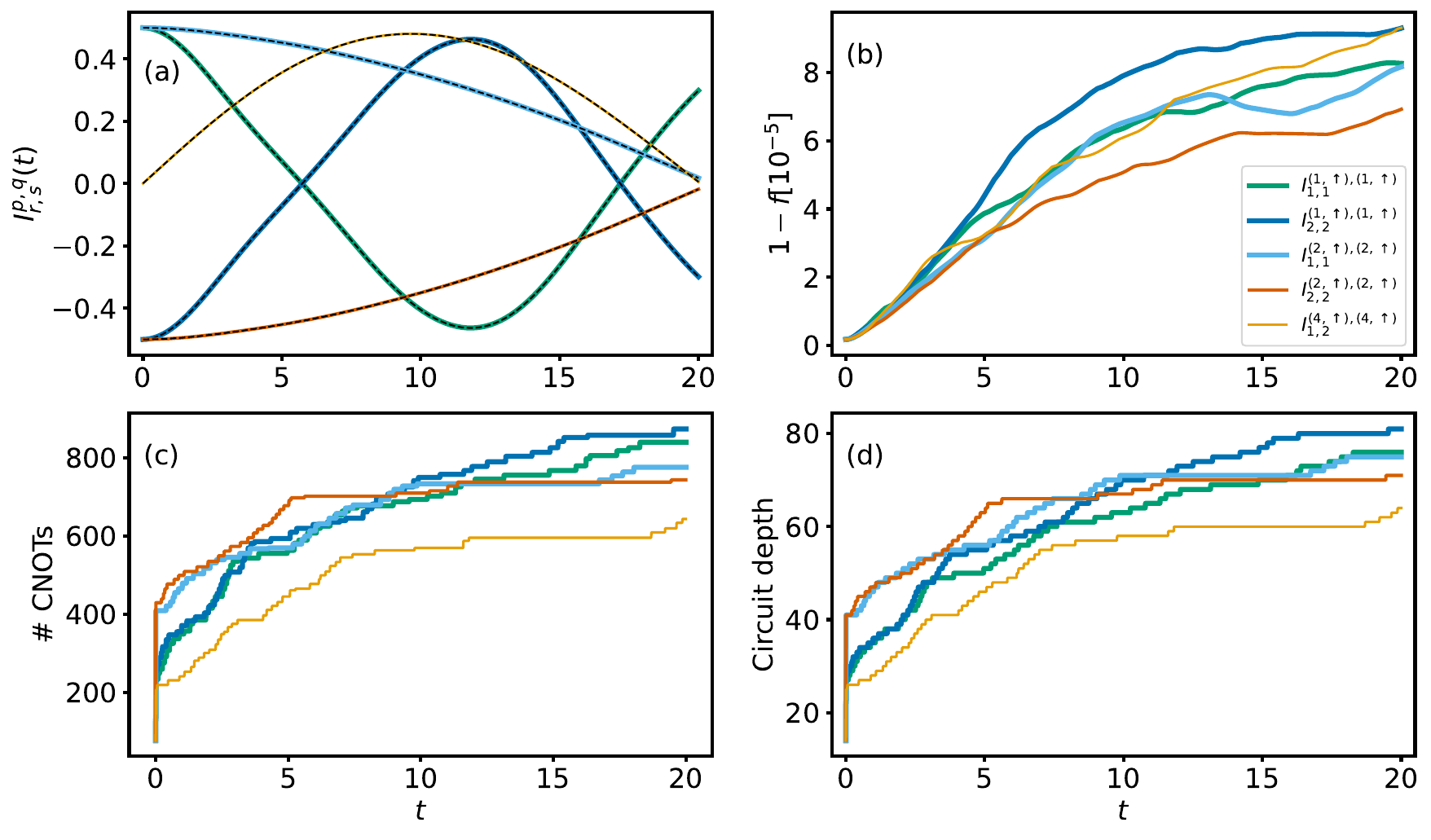}
				\caption{Numerical simulation of the AVQDS approach for computing the single-particle Green's function of the molecule LiH. (a) Examples of $I^{p,q}_{\alpha,\beta}(t)$ dynamics for five different combinations of $p$, $q$, $\alpha$, and $\beta$ (solid lines), obtained by evaluating the AVQDS quantum circuit in Figure~\figref{fig1}{b}. The exact simulations (black dashed lines) via exact diagonalization (eq~\ref{eq:ED}) are also shown for comparison. (b) The corresponding infidelities $1-f$ illustrate the high accuracy of AVQDS in calculating the Green's function components $I^{p,q}_{\alpha,\beta}(t)$, with an infidelity below $9.3\times 10^{-5}$ during the studied time window. (c) The corresponding number of CNOT gates, which increases from an initial count of 78 to a maximum of 874 at the final simulation time of $t=20$. (d) The corresponding circuit depth, which increases from 14 to a maximum circuit depth of 81 at $t=20$.}
		\label{fig8} 
\end{center}
\end{figure*}

Figure~\figref{fig8}{a} presents examples of the simulated $I^{p,q}_{\alpha,\beta}$ dynamics for five different components, where calculations using the adaptive variational quantum circuit (solid lines) presented in Figure~\figref{fig1}{b} agree very well with the reference results (black dashed lines) from exact diagonalization. The corresponding infidelity in Figure~\figref{fig8}{b} stays below $9.3\times 10^{-5}$ within the studied time window which confirms that AVQDS can be used to accurately calculate the real-time Green's function components.

The required near-term quantum resources in terms of CNOT gates are plotted in Figure~\figref{fig8}{c} for the $I^{p,q}_{\alpha,\beta}$ simulations presented in Figure~\figref{fig8}{a}. AVQITE prepares the ground state with an infidelity of $1-f = 1.8\times 10^{-6}$. The corresponding variational ansatz includes 6 two- and 11 four-qubit rotation gates, yielding 78 CNOT gates assuming full qubit connectivity. The number of CNOTs initially increases rapidly during the time evolution, and tends to saturate at later times. The maximal number of CNOT gates at the final simulation time is in the range of [640, 874] across all the components. The corresponding dynamics of the circuit depth is plotted in Figure~\figref{fig8}{d}. The depth of the ground state circuit is 14 and grows to 81 at the final simulation time of $t=20$.

\begin{figure*}[t!]
\begin{center}
		\includegraphics[scale=0.55]{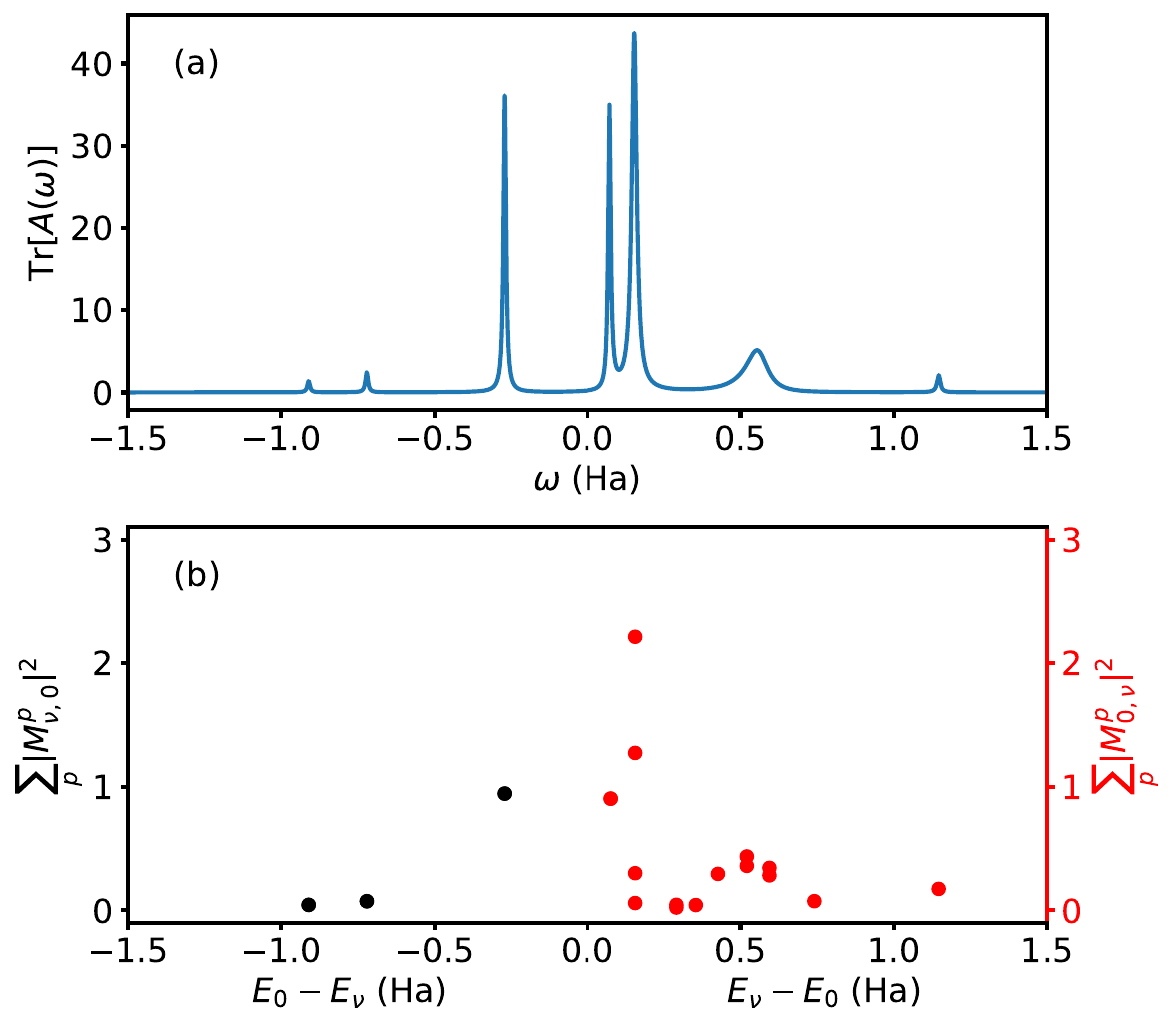}
\caption{Spectral function of LiH. (a) Trace of spectral function $\mathrm{Tr}\left[A(\omega)\right]$ calculated with compressive sensing, which shows several discrete delta peaks. (b) $\sum_p|M^p_{\nu,0}|^2$ (black dots) and $\sum_p|M^p_{0,\nu}|^2$ (red dots) as a function of the energy differences $E_0-E_\nu$ and $E_\nu - E_0$, respectively. Only the dominant transition amplitudes with $\sum_p|M^p_{\nu,\mu}|^2 > 0.02$ are shown. The peaks in the spectral functions originate from transitions between the ground state with energy $E_0$ to the excited states with energies $E_\nu$, and vice versa.}
		\label{fig9} 
\end{center}
\end{figure*}

We next present the trace of the spectral function of LiH, $\mathrm{Tr}\left[A(\omega)\right] = -\frac{1}{\pi}\mathrm{Tr}\left[\mathrm{Im}[G^\mathrm{R}(\omega)]\right]$, which represents the total density of states. Here, $G^\mathrm{R}(\omega) = \int_{-\infty}^{\infty}\mathrm{d}t\,\mathrm{e}^{\mathrm{i}(\omega+\mathrm{i}\,\varepsilon)t}G^\mathrm{R}(t)$ is a $10 \times 10$ (including spin degeneracy) matrix at each $\omega$ point. Figure~\figref{fig9}{a} shows the result obtained with compressive sensing without broadening in the Fourier transformation. For comparison, in Figure~\figref{fig9}{b} we also show $\sum_p|M^p_{\nu,0}|^2$ (black dots) and $\sum_p|M^p_{0,\nu}|^2$ (red dots) as a function of the energy differences $E_0 - E_\nu$ and $E_\nu - E_0$, respectively. The spectral function shows all dominant transitions present in Figure~\figref{fig9}{b}. Specifically, $\mathrm{Tr}\left[A(\omega)\right]$ exhibits three dominant peaks. The peak around $\omega\approx -0.27$ Ha in the spectral function in Figure~\figref{fig9}{a} results from transitions between the ground state and doubly degenerate excited states with spin $S=1/2$ and electron number of $N_e=1$. For comparison, the ground state has $S=0$ while $N_e=2$ due to the treatment of the Li $1s$ orbital as a frozen core orbital. The dominant signals at positive frequencies, $\omega\approx 0.08$ Ha and $\omega\approx 0.16$~Ha, in Figure~\figref{fig9}{a} arise from transitions between the ground state and doubly degenerate excited states with spin $S=1/2$ but $N_e=3$. As a result, the spectral function provides direct insights into the excitation levels of the molecular system similar to the study of the Fermi-Hubbard models in section~\ref{sec:GF_res}.

\subsection{\label{sec:chi3} Third-order susceptibility of two-site quantum spin-1 model}

\subsubsection*{Model}

To demonstrate the calculation of the third-order susceptibility using the CUL circuit presented in section~\ref{sec:chi}, we investigate a higher-spin model that has been used to describe 2DCS experiments on rare-earth orthoferrites~\cite{mootz2023twodimensional}. The Hamiltonian is given by:
\begin{align}
\label{eq:Ham_spin}
\hat{\mathcal{H}} =&\,J\sum_{i=1}^{N-1}\hat{\mathbf{S}}_i\cdot\hat{\mathbf{S}}_{i+1}-\mathbf{D}\cdot\sum_{i=1}^{N-1}\hat{\mathbf{S}}_i\times\hat{\mathbf{S}}_{i+1}\,.
\end{align}
The first term of the Hamiltonian characterizes the antiferromagnetic coupling between nearest neighbors with an exchange constant $J > 0$. The second term in eq~\ref{eq:Ham_spin} describes the Dzyaloshinskii-Moriya (DM) interaction with an antisymmetric exchange vector $\mathbf{D}$, which we assume to be aligned along the $y$-direction, i.e., $\mathbf{D}=D\mathbf{y}$. 

\begin{figure}[t!]
\begin{center}
		\includegraphics[scale=0.45]{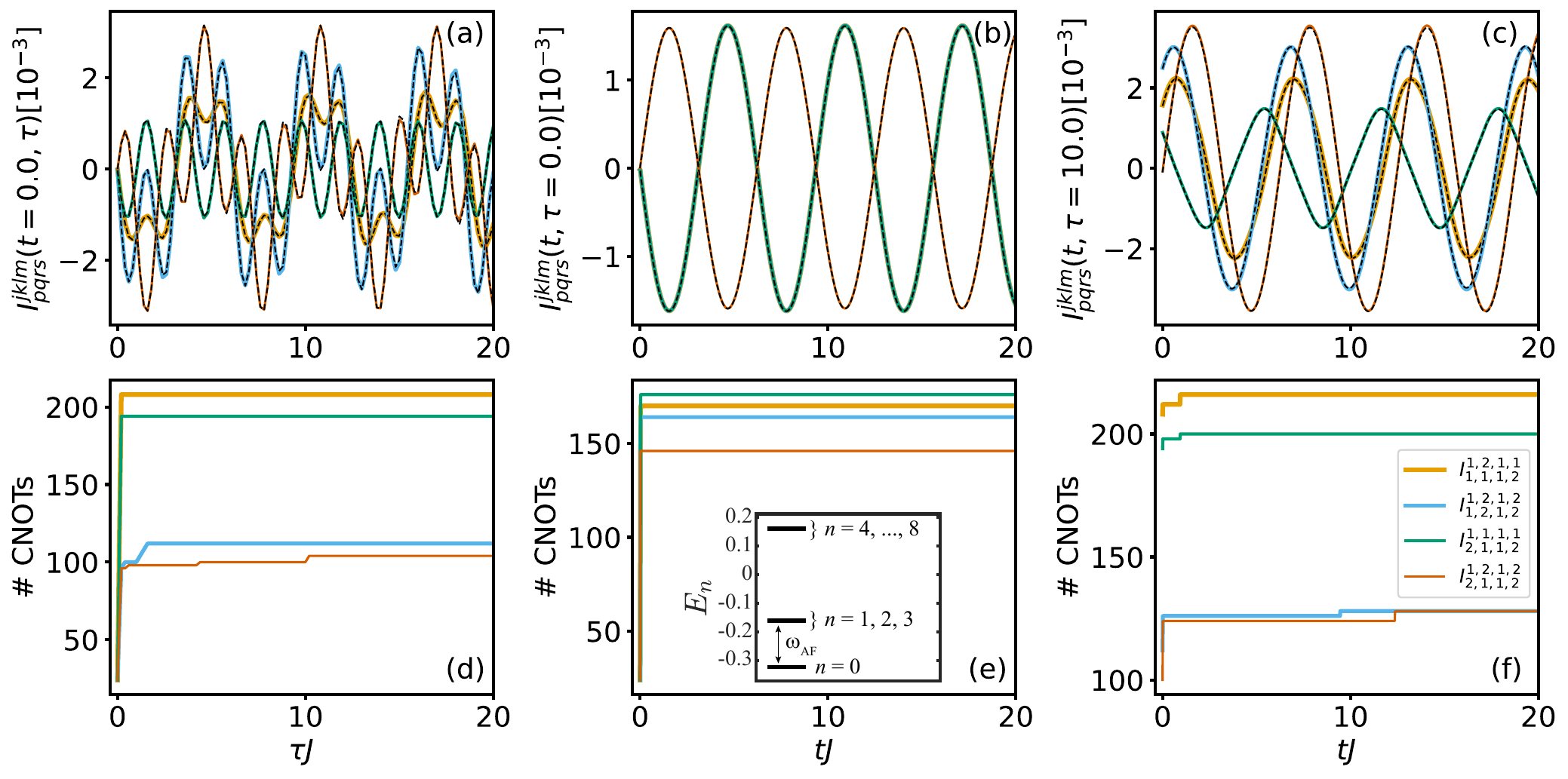}
		\caption{Numerical simulation of the AVQDS algorithm for calculating the third-order nonlinear susceptibility of the two-site spin-1 model. (a)--(c) Four examples of $I^{jklm}_{pqrs}(t,\tau)$ (a) as a function of $\tau$ at fixed $t J=0$, and as a function of time $t$ at fixed (b) $\tau J=0$ and (c) $\tau J=10$.  The dynamics obtained by the AVQDS approach (solid lines) accurately reproduce the exact dynamics (dashed black lines) for all presented $I^{jklm}_{pqrs}$. (d)--(f) The corresponding dynamics of the required number of CNOT gates. The number of CNOT gates significantly increases only at earlier times and saturates after a time of about $\tau J=1.5$ in (d) and $t J=1.3$ in (e) and (f). The saturated number of CNOTs at the final simulation time of $t J=20$ falls within the range of 128 to 216. Inset of (e): Eigenenergies $E_n$ of the quantum spin Hamiltonian in eq~\ref{eq:Ham_spin}. The Hamiltonian exhibits $(2s+1)^N=9$ eigenstates for $N=2$ sites and spin-$s=1$. The magnon frequency $\omega_\mathrm{AF}=E_1-E_0\approx 0.16$ is given by the energy difference between the ground state ($n = 0$) and the first excited state ($n = 1$). The states $n=1,2,3$ as well as $n=4,\cdots,8$ are nearly degenerate.}
		\label{fig5} 
\end{center}
\end{figure}

In the simulations, we consider a two-site spin-1 model for demonstration. Note that the quantum resource scaling with respect to system size and spin magnitude $s$ for the ground state and dynamics simulations of this model have been numerically studied in refs~\citenum{mootz2023twodimensional, getelina2024}. We define the energy unit by setting the coupling constant $J$ to one, while using a Dzyaloshinskii-Moriya interaction strength of $D=0.2$. The inset of Figure~\ref{fig5}(e) shows the energy levels $E_n$ of the analyzed two-site spin-1 model obtained via exact diagonalization. The energy difference between the ground state ($n=0$) and the first excited state ($n=1$) determines the magnon frequency $\omega_\mathrm{AF}=E_1-E_0\approx 0.16$, represented by the double arrow. Notably, the eigenenergies of states $n=1,2,3$ as well as the states $n=4,\cdots, 8$ are nearly degenerate and the energy difference between $E_{n=1,2,3}$ and $E_{n=4,\cdots, 8}$ is approximately $2\omega_\mathrm{AF}$ (see inset of Figure~\ref{fig5}), indicating a quasi-harmonic energy spectrum of the two-site spin-1 model, as discussed in more detail in ref~\citenum{mootz2023twodimensional}. 

To map the spin-1 operators to multi-qubit operators, we use the Gray code~\cite{matteo2021} which provides shallower quantum circuits compared to the binary encoding in calculating two-dimensional coherent spectra, as demonstrated in ref~\citenum{mootz2023twodimensional}. The encoding of the spin-1 operators requires $n_q=2$ qubits for a single spin such that the Hamiltonian eq~\ref{eq:Ham_spin} for $N=2$ sites is represented by $N_q=n_q N=4$ qubits. 

\subsubsection*{Ground state preparation}

To prepare the ground state $\ket{\mathrm{G}}$ of the quantum-spin Hamiltonian eq~\ref{eq:Ham_spin}, we employ the AVQITE method~\cite{AVQITE}. 
We utilize $\ket{\varphi_0} = \ket{0}^{\otimes N\mathrm{q}}$ as our reference state. For the ground state preparation, we adopt the following operator pool: 
\begin{align}
\mathcal{P}=\{\sigma^y_i\}_{i=1}^{N_q}\cup \{\sigma^y_{i} \sigma^z_{i+1}\}_{i=1}^{N_q-1}\cup \{\sigma^z_{i} \sigma^y_{i+1}\}_{i=1}^{N_q-1}\,.
\label{eq:pool0}
\end{align}
For the third-order susceptibility calculation with AVQDS, we utilize the following pool:
\begin{align}
    \mathcal{P}=&\{A_i: A \in \{\sigma^x,\sigma^y,\sigma^z\},\, 1 \le i \le N_\mathrm{q} \}  \nonumber \\
    &\cup \{A_i B_j: A,B \in \{\sigma^x,\sigma^y,\sigma^z\},\, 1\le i < j \le N_\mathrm{q}\}\,. 
    \label{eq:pool}
\end{align}
This pool contains all possible one- and two-qubit Pauli words.

\subsubsection*{Simulation results}

To demonstrate the calculation of the third-order susceptibility using the CUL circuits with AVQDS presented in section~\ref{sec:chi_form}, we focus on the third-order susceptibility $\chi_{zzzz}^{(3)}(t,\tau,0)$. This susceptibility is particularly relevant for analyzing 2DCS experiments employing a collinear two-pulse geometry, where the applied magnetic field consists of two copropagating pulses polarized along the $z$-direction. In such a scenario, the third and fourth terms within the square brackets in eq~\ref{eq:chi3_trans} become identical. Since the transformation eq~\ref{eq:s_trafo} involves $n_z=2$ terms for $\hat{S}^z_j$ and spin $s=1$~\cite{mootz2023twodimensional}, the summation in eq~\ref{eq:chi3_trans} encompasses $3 N^4 n_z^4=768$ terms, each of which needs to be calculated using the CUL circuit depicted in Figure~\figref{fig4}{b}.

To illustrate the calculation of $\chi_{zzzz}^{(3)}(t,\tau,0)$ with the CUL circuit in Figure~\figref{fig4}{b}, we use the first term within the square brackets of eq~\ref{eq:chi3_trans} as an example:
\begin{align}
 &I^{jklm}_{pqrs}(t,\tau) \nonumber \\
&\equiv\mathrm{Im}[\langle \mathrm{e}^{\mathrm{i}\hat{\mathcal{H}}(t+\tau)}P^z_{j,p} \mathrm{e}^{-\mathrm{i}\hat{\mathcal{H}}t}P^z_{k,q} \mathrm{e}^{-\mathrm{i}\hat{\mathcal{H}}\tau} P^z_{l,r}P^z_{m,s}\rangle] \nonumber   \\
&\approx \mathrm{Im}[\bra{G[\bth^1]} U_\tau^\dagger(\bth^2)U_t^\dagger(\bth^3) P^z_{j,p} U_t(\bth^3)P^z_{k,q} \nonumber \\
& \qquad\qquad\quad\times U_\tau(\bth^2) P^z_{l,r}P^z_{m,s}\ket{G[\bth^1]}]\,.
\end{align}
We first evolve the state in eq~\ref{eq:qc_chi_1} with $P_0=P_1 =I^{\otimes N_q}$, $P_4=P^z_{l,r}$, and $P_5=P^z_{m,s}$ up to time $\tau$ using AVQDS. After applying the $X$-gate and controlled $P_3=P^z_{k,q}$ gate, we further propagate the state in eq~\ref{eq:qc_chi_2} to $t\in [0,t_\mathrm{max}]$, for which we adopt a uniform time mesh. Similar circuit simulations are repeated for $\tau \in [0,\tau_\mathrm{max}]$ with a uniform mesh.

Figure~\figref{fig5}{a} shows four examples of $I^{jklm}_{pqrs}(t,\tau)$ as a function of $\tau$ at fixed $t=0$, while Figure~\figref{fig5}{b} and \figref{fig5}{c} present the corresponding $I^{jklm}_{pqrs}(t,\tau)$ as a function of $t$ at fixed $\tau=0$ and $\tau J=10$, respectively.   
The results obtained from the AVQDS approach (solid lines) are compared with the exact simulation results obtained by evolving the states in eqs~\ref{eq:qc_chi_1} and \ref{eq:qc_chi_2} using exact diagonalization eq~\ref{eq:ED} (dashed black lines). The results from the CUL circuit simulations agree with the exact dynamics for all presented $I^{jklm}_{pqrs}$, with fidelities exceeding $99.99\%$. 

The corresponding number of CNOT gates as a function of time are shown in Figure~\figref{fig5}{d}--\figref{fig5}{f}. The initial $22$ CNOTs at $t=\tau=0$ are determined by the ground state ansatz, which is obtained using AVQITE with an infidelity of about $10^{-10}$. Throughout the time evolution, the number of CNOT gates only increases rapidly
at very early times, followed by saturation after a time of about $\tau J=1.5$ (Figure~\figref{fig5}{d}) and $t J=1.3$ (Figure~\figref{fig5}{e} and \figref{fig5}{f}). The saturated number of CNOTs at the final simulation time of $t J=20$ ranges from 128 to 216 for the various $I^{jklm}_{pqrs}(t,\tau)$.

\begin{figure}[t!]
\begin{center}
		\includegraphics[scale=0.65]{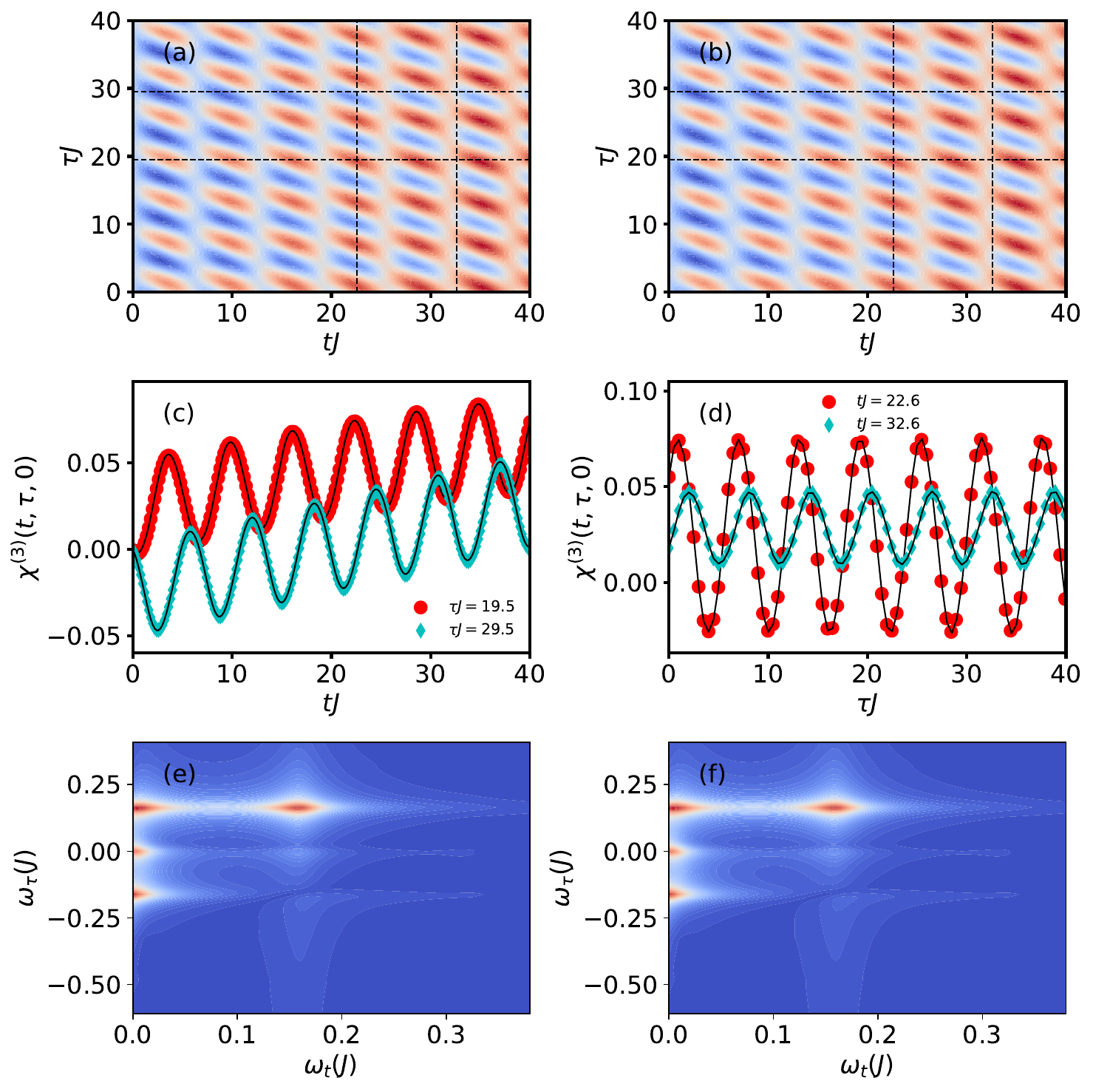}
		\caption{Third-order susceptibility of
the two-site spin-1 system in the two-dimensional time and frequency domains. (a) $\chi^{(3)}_{zzzz}(t,\tau,0)$  as a function of times $t$ and $\tau$ calculated using the CUL circuit, which agrees with the exact result in (b) obtained by evaluating eq~\ref{eq:chi3ex}. Slices shown in (c) and (d) are indicated by horizontal and vertical dashed lines. (c), (d) Slices of $\chi^{(3)}_{zzzz}(t,\tau,0)$ from (a) and (b) at two fixed times $\tau$ (c) and $t$ (d). The AVQDS results (circles) accurately reproduce the exact dynamics (solid black lines).  (e), (f)  Two-dimensional (2D) Fourier
transform of $\chi^{(3)}_{zzzz}(t,\tau,0)$ from (a) and (b), respectively. Both 2D spectra are in excellent agreement.}
		\label{fig6} 
\end{center}
\end{figure}

To obtain the third-order susceptibility $\chi^{(3)}_{zzzz}(t,\tau,0)$, we simulated the different contributions in eq~\ref{eq:chi3_trans} using the CUL circuit in Figure~\figref{fig4}{b}. Simulations were performed up to $\tau_\textrm{max} J=40$ with a step size of $\delta\tau=0.5$, which provides sufficient resolution of the signals in the 2D spectrum of the third-order susceptibility. To calculate $\chi^{(3)}_{zzzz}(t,\tau,0)$, we interpolated the different contributions in eq~\ref{eq:chi3_trans} to a $401 \times 401$ uniform mesh of $t$ and $\tau$ with $t_\textrm{max}J = \tau_\textrm{max}J = 40$ before performing the summation over the components in eq~\ref{eq:chi3_trans}. Figure~\figref{fig6}{a} shows $\chi^{(3)}_{zzzz}(t,\tau,0)$ as a function of times $t$ and $\tau$ from statevector simulations of the CUL circuits. The corresponding result for the exact dynamics, derived in Appendix~\ref{sec:chi3app}, is presented in Figure~\figref{fig6}{b}. The third-order susceptibility obtained by the CUL circuits agrees well with the exact simulation result. This is further demonstrated by the slices at fixed $\tau$ and fixed $t$ plotted in Figure~\figref{fig6}{c} and \figref{fig6}{d}, where the CUL circuit results (circles) match the exact result (solid black line) over the full range of simulation times.

Figure~\figref{fig6}{e} and \figref{fig6}{f} show the resulting 2D spectra of the third-order susceptibility, $\chi^{(3)}_{zzzz}(\omega_t,\omega_\tau,0)$, obtained after performing the 2D discrete Fourier transformation of $\chi^{(3)}_{zzzz}(t,\tau,0)$ from Figure\figref{fig6}{a} and \figref{fig6}{b} using the Pad\'e approximation in both the $t$ and $\tau$-directions. The 2D spectrum of the third-order susceptibility transformed from the real-time CUL circuit results agrees well with that from exact diagonalization. The 2D spectrum exhibits four dominant peaks at $(\omega_t,\omega_\tau)=(0,\pm\omega_\mathrm{AF})$, $(\omega_t,\omega_\tau)=(0,0)$, and $(\omega_t,\omega_\tau)=(\omega_\mathrm{AF},\omega_\mathrm{AF})$. As demonstrated in ref~\citenum{mootz2023twodimensional} and discussed in Appendix~\ref{sec:chi3app}, these signals originate from transitions between the ground state $|\Psi_0\rangle$ and the first excited state $|\Psi_1\rangle$, as well as between $|\Psi_1\rangle$ and the third excited state $|\Psi_3\rangle$. Such signals directly manifest in the 2DCS spectra, underscoring the significance of nonlinear susceptibilities in interpreting 2DCS experiments.

\section{\label{sec:con} Conclusion and outlook}

In this work, we presented and benchmarked quantum computing methods to calculate single-particle Green’s functions and nonlinear susceptibilities, adopting controlled-unitaries-liberated circuits and adaptive variational algorithms for ground state preparation and real-time propagation. For the computation of single-particle Green’s functions, we followed the CUL approach outlined in ref~\citenum{endo2020calculation}, utilizing AVQITE to generate ground state circuits and AVQDS to approximate the time-evolution operator instead of CUL with HVA as in ref~\citenum{endo2020calculation}. To illustrate the CUL approach, we computed the single-particle Green’s function of Fermi-Hubbard chains with $N=4$ and $N=6$ sites. Specifically, we evaluated the real-time Green’s function in momentum space as well as its corresponding spectral function, and compared the results with exact diagonalization calculations using the Lehmann representation of the Green’s function. Our findings demonstrate that the CUL approach with AVQDS can accurately simulate the dynamics of Green’s functions over sufficiently long times and obtain reliable spectral functions.
The CUL quantum circuits required for the simulation of the Green’s functions are shallower compared to those obtained with HVA in ref~\citenum{endo2020calculation}. Furthermore, in comparison to the calculation of the single-particle Green's function with the CUR method using AVQDS in ref~\citenum{Gomes2023}, our results demonstrate that utilizing the CUL approach with AVQDS can reduce the circuit complexity by bypassing the state overlap measurement. Comparable CNOT counts but deeper quantum circuits are required for CUR with HVA~\cite{Libbi2022}. 
As another example to demonstrate the general applicability of the CUL approach with AVQDS, we applied it to accurately evaluate the Green's function of the molecule LiH, which amounts to an estimated circuit depth up to 81 layers.

We also extended the CUL method from calculating single-particle Green's functions to evaluating nonlinear susceptibilities, which is crucial to explain the two-dimensional coherent spectroscopy experiments on semiconductors~\cite{Kuehn2009, Kuehn2011, Junginger2012}, superconductors~\cite{Higgs_2dTHz,Luo2019}, topological systems~\cite{Bhandia2024}, quantum spin systems~\cite{Huang2024,mootz2023twodimensional}, and molecular systems~\cite{Ghalgaoui2020,Zhang2021}. Here, the nonlinear susceptibilities depend on two times such that their computation via the quantum circuit presented in Figure~\figref{fig4}{b} requires the application of the AVQDS algorithm for two times. To demonstrate the validity of our method, we studied an antiferromagnetic quantum spin model including a Dzyaloshinskii-Moriya interaction. We calculated the third-order nonlinear susceptibility in the 2D time and frequency domains for a two-site spin-1 model and compared the results with numerical exact data. The third-order susceptibility calculated using the quantum computing approach agrees well with the exact result which confirms the accuracy of the CUL method in evaluating higher-order correlation functions using AVQDS for state evolution.

Several modifications of the presented approach are feasible, where the quantum circuit complexity can potentially be reduced by trading for more measurements. For example, the Hadamard test circuits could be replaced with direct measurement circuits as shown in ref~\citenum{mitarai2019methodology}. An interesting direction for future work is motivated by the quantum algorithm proposed in ref~\citenum{Kemper2024}, which was developed to study long-time dynamics and correlation functions of driven-dissipative quantum systems, where the steady state of the time evolution is stable to perturbations. Compared with the standard Hadamard test circuit where an ancilla qubit is required to maintain entanglement and coherence with the system qubits during the entire simulation time, the key idea is to set the ancilla to the $\ket{+}$ state before applying the ancilla-controlled Pauli gates and measure the ancilla immediately after the ancilla-system entangling gates at each of the specified times to decouple the ancilla from the system qubits. The $n$-point correlation function can then be constructed based on the measurement outcomes. The ancilla qubit is only entangled with the system qubits for a minimal amount of time when the ancilla-system entangling gates are applied. This promises a great advantage when the bottleneck for successful execution of the measurement circuits is the coherence requirement of the ancilla qubit, like the steady state of a driven-dissipative quantum system. For simulations of isolated quantum systems as in this work, no clear advantage is expected due to the stringent coherence requirement for both ancilla and system qubits~\citenum{Kemper2024}. However, a potential advantage may still be achieved thanks to the adaptive variational quantum algorithms, which generate highly compressed quantum circuits for dynamics simulations and substantially reduce the coherence time requirement. By adopting mid-circuit measurements for the ancilla qubit which disentangles the ancilla from the system qubits prior to time evolution, the state propagation only needs to be applied to the system qubits, which can potentially simplify the AVQDS circuits.

The presented algorithms for computing single-particle Green’s functions and nonlinear susceptibilities can readily be extended to calculate higher-order multi-time correlation functions that depend on more than two times. Here, it would be interesting to compare the quantum resources required by the CUL method used in this paper against those required by the CUR approach considered in refs~\citenum{Libbi2022,Gomes2023}. In addition, the approach can be naturally extended to finite temperature calculations by adopting techniques such as a quantum version of the minimally entangled typical thermal states algorithm~\citenum{getelina2023adaptive, qite_chan20, White2009MinimallyET}.

Regarding the practical implementation of the AVQDS approach for computing high-order correlation functions on quantum hardware, it is essential to investigate the impact of hardware noise and shot noise resulting from a finite number of measurements. Here, the incorporation of error mitigation methods plays a crucial role~\cite{caiQuantumErrorMitigation2022, Getelina2024QuantumSE}.

\appendix

\section{\label{sec:G_qc} Quantum circuit for calculating single-particle Green's functions}

In this appendix, we demonstrate that the quantum circuit in Figure~\figref{fig1}{b} evaluates eq~\ref{eq:Ipqv}. The system qubits are initialized in a reference product state $\ket{\varphi_0}$, which can be straightforwardly prepared on a quantum computer. Then, the ground state is prepared by applying the unitary operator $U_\mathrm{G}[\bth^1]$ following AVQITE, yielding $\ket{G[\bth^1]}=U_\mathrm{G}[\bth^1]\ket{\varphi_0}$. The initial state of the ancilla qubit is $\ket{0}$, which becomes $(\ket{0}+\ket{1})/\sqrt{2}$ after applying the Hadamard gate. The quantum state after applying $U_\mathrm{G}$ on the system qubits and $H$ on the ancilla qubit reads: 
\begin{align}
    \ket{\Psi}=\frac{1}{\sqrt{2}}\left[\ket{0}\otimes\ket{G[\bth^1]}+\ket{1}\otimes\ket{G[\bth^1]}\right]\,.
\end{align}
After applying the controlled $P_\beta$ operation controlled by the ancilla qubit, the state is given by $\frac{1}{\sqrt{2}}\ket{0}\otimes\ket{G}+\frac{1}{\sqrt{2}}\ket{1}\otimes P_\beta\ket{G}$. The time-evolving state under AVQDS with time-dependent parameters $\bth^1(t)$ and $\bth^2(t)$ becomes:
\begin{align}
    \ket{\Psi}=&\frac{1}{\sqrt{2}}\ket{0}\otimes U_t[\bth^2]\ket{G[\bth^1]} \\ \nonumber 
    &+\frac{1}{\sqrt{2}}\ket{1}\otimes U_t[\bth^2] P_\beta\ket{G[\bth^1]}\,.
\end{align}  
Applying the $X$-gate on the ancilla qubit followed by the controlled $P_\alpha$ operation yields 
\begin{align}
    \ket{\Psi}=&\frac{1}{\sqrt{2}}\ket{1}\otimes P_\alpha U_t[\bth^2]\ket{G[\bth^1]} \\ \nonumber
    &+\frac{1}{\sqrt{2}}\ket{0}\otimes U_t[\bth^2] P_\beta\ket{G[\bth^1]}\,.
\end{align}
After the application of the Hadamard gate on the ancilla, the quantum state is given by
\begin{align}
    \ket{\Psi}&=\frac{1}{2}\ket{0}\otimes\left(P_\alpha U_t[\bth^2]+U_t[\bth^2]P_\beta\right)\ket{G[\bth^1]} \nonumber \\
    &-\frac{1}{2}\ket{1}\otimes\left(P_\alpha U_t[\bth^2]-U_t[\bth^2]P_\beta\right)\ket{G[\bth^1]}\,.
\end{align}
Finally, performing a $Z$ measurement on the ancilla qubit, $\bra{\Psi}\hat{\sigma}^z\otimes I^{\otimes N_q}\ket{\Psi}$, yields $I^{p,q}_{\alpha,\beta} = 2p_{\ket{0}}-1$ as defined in eq~\ref{eq:Ipqv}, where $p_{\ket{0}}$ is the probability that the ancilla is measured to be in the state $\ket{0}$.

\section{\label{sec:Lehmann} Lehmann representation of spectral function}

In this appendix, we derive the Lehmann representation of the spectral function. Using the completeness relation $\hat{I}=\sum_\nu\ket{\Psi_\nu}\bra{\Psi_\nu}$, where $\ket{\Psi_\nu}$ are the eigenstates of the fermionic Hamiltonian $\h$ with energies $E_\nu$, eq~\ref{eq:Gr} can be written as
\begin{align}
    G^\mathrm{R}_{p,q}(t)=-\mathrm{i}\,\Theta(t)&\left[\mathrm{e}^{\mathrm{i} (E_0-E_\nu) t}T^p_{0,\nu}\left(T^q_{0,\nu}\right)^\star\right.\nonumber \\
    &\left. \mathrm{e}^{\mathrm{i} (E_0-E_\nu) t}T^p_{\nu,0}\left(T^q_{\nu, 0}\right)^\star\right]\,,
\end{align}
with transition matrix elements $T^p_{\mu,\nu}\equiv\bra{\Psi_\mu}\ca_p\ket{\Psi_\nu}$.
By applying the Fourier transform
\begin{align}
    G^\mathrm{R}_{p,q}(\omega)=\int_{-\infty}^{\infty}\mathrm{d}t\,\mathrm{e}^{\mathrm{i}(\omega+\mathrm{i}\,\varepsilon)t}G^\mathrm{R}_{p,q}(t)
\end{align}
with an infinitesimal $\varepsilon$ to guarantee convergence of the integral, we find
\begin{align}
    G^\mathrm{R}_{p,q}(\omega)=\sum_\nu\left[\frac{T^p_{0,\nu}\left(T^q_{0,\nu}\right)^\star}{E_0-E_\nu+\omega+\mathrm{i}\,\varepsilon}+\frac{T^p_{\nu,0}\left(T^q_{\nu,0}\right)^\star}{E_\nu-E_0+\omega+\mathrm{i}\,\varepsilon}\right]\,.
\end{align}
Finally, the retarded Green's function in momentum space is given by
\begin{align}
\label{eq:GRk}
    G^\mathrm{R}_{k}&=\frac{1}{N}\sum_{p,q}G^\mathrm{R}_{p,q}\mathrm{e}^{-\mathrm{i}\,k(p-q)} \nonumber \\
    &=\frac{1}{N}\sum_{p,q}\sum_\nu\left[\frac{T^p_{0,\nu}\left(T^q_{0,\nu}\right)^\star}{E_0-E_\nu+\omega+\mathrm{i}\,\varepsilon}\right. \nonumber \\
    &\left.+\frac{T^p_{\nu,0}\left(T^q_{\nu,0}\right)^\star}{E_\nu-E_0+\omega+i\varepsilon}\right]\mathrm{e}^{-\mathrm{i}\,k(p-q)}\,,
\end{align}
which simplifies to eq~\ref{eq:GL} for $k=0$.

\section{\label{sec:chi_qc} Quantum circuit for calculating the third-order susceptibility}

In this appendix, we demonstrate that the quantum circuit presented in Figure~\figref{fig4}{b} yields the imaginary part of the expectation value 
$\bra{G[\bth^1]} P_0 P_1 U_\tau^\dagger[\bth^3] U_t^\dagger[\bth^2] P_2\, U_t[\bth^2] P_3\, U_\tau[\bth^3] P_4 P_5\ket{G[\bth^1]}$. The system is initialized in a reference product state $\ket{\varphi_0}$. The ground state is prepared by applying the unitary operator $U_\mathrm{G}[\bth^1]$ using AVQITE, leading to $\ket{G[\bth^1]}=U_\mathrm{G}[\bth^1]\ket{\varphi_0}$. The ancilla qubit, initially in the state $\ket{0}$, is given by $2^{-1/2}(\ket{0}+\mathrm{i}\ket{1})$ after applying the Hadamard gate $H$ and $S$-gate. As a result, the quantum state before applying the controlled $P_5$ gate reads
\begin{align}
    \ket{\Psi}=\frac{1}{\sqrt{2}}\left[\ket{0}\otimes\ket{G[\bth^1]}+\mathrm{i}\ket{1}\otimes\ket{G[\bth^1]}\right]\,.
\end{align}
Applying the controlled $P_5$ and $P_4$ operations controlled by the ancilla qubit leads to the quantum state
\begin{align}
    \ket{\Psi}=\frac{1}{\sqrt{2}}\left[\ket{0}\otimes\ket{G[\bth^1]}+\mathrm{i}\ket{1}\otimes P_4 P_5\ket{G[\bth^1]}\right]\,.
\end{align}
Applying the $X$ gate on the ancilla qubit followed by the controlled $P_0$ and $P_1$ gates yields
\begin{align}
    \ket{\Psi}=\frac{1}{\sqrt{2}}\left[\ket{1}\otimes P_1 P_0\ket{G[\bth^1]}+\mathrm{i}\ket{0}\otimes P_4 P_5\ket{G[\bth^1]}\right]\,.
\end{align}
The time-evolving state under AVQDS with time-dependent variational parameters $\bth^1(\tau)$ and $\bth^3(\tau)$ becomes
\begin{align}
    \ket{\Psi}=\frac{1}{\sqrt{2}}&\left[\ket{1}\otimes U_\tau[\bth^3]P_1 P_0\ket{G[\bth^1]}\right. \nonumber \\
    &\left.+\mathrm{i}\ket{0}\otimes U_\tau[\bth^3]P_4 P_5\ket{G[\bth^1]}\right]\,.
 \label{eq:qc_1}   
\end{align}
Application of the $X$ gate followed by the controlled $P_3$ gate leads to
\begin{align}
    \ket{\Psi}=\frac{1}{\sqrt{2}}&\left[\ket{0}\otimes U_\tau[\bth^3]P_1 P_0\ket{G[\bth^1]}\right. \nonumber \\
    &\left.+\mathrm{i}\ket{1}\otimes P_3 U_\tau[\bth^3]P_4 P_5\ket{G[\bth^1]}\right]\,.
\end{align}
Propagating this state in time using AVQDS with time-dependent variational parameters $\bth^1(t)$, $\bth^3(t)$, and $\bth^2(t)$ produces
\begin{align}
    \ket{\Psi}=\frac{1}{\sqrt{2}}&\left[\ket{0}\otimes U_t[\bth^2]U_\tau[\bth^3]P_1 P_0\ket{G[\bth^1]}\right. \nonumber \\
    &\left.+\mathrm{i}\ket{1}\otimes U_t[\bth^2]P_3 U_\tau[\bth^3]P_4 P_5\ket{G[\bth^1]}\right]\,.
\end{align}
The application of the $X$ gate followed by the controlled $P_2$ gate results in the state
\begin{align}
    \ket{\Psi}=\frac{1}{\sqrt{2}}&\left[\ket{1}\otimes P_2 U_t[\bth^2]U_\tau[\bth^3]P_1 P_0\ket{G[\bth^1]}\right. \nonumber \\
    &\left.+\mathrm{i}\ket{0}\otimes U_t[\bth^2]P_3 U_\tau[\bth^3]P_4 P_5\ket{G[\bth^1]}\right]\,.
\end{align}
After executing the Hadamard gate on the ancilla qubit, the quantum state is given by
\begin{align}
    \ket{\Psi}&=\frac{1}{2}\ket{0}\otimes\left[P_2 U_t[\bth^2]U_\tau[\bth^3]P_1 P_0 \right. \nonumber \\
    &\qquad\qquad\left.+i U_t[\bth^2]P_3 U_\tau[\bth^3]P_4 P_5\right]\ket{G[\bth^1]} \nonumber \\
    &-\frac{1}{2}\ket{1}\otimes\left[P_2 U_t[\bth^2]U_\tau[\bth^3]P_1 P_0 \right. \nonumber \\
    &\qquad\qquad\left.-i U_t[\bth^2]P_3 U_\tau[\bth^3]P_4 P_5\right]\ket{G[\bth^1]}\,.
\end{align}
Performing a $Z$ measurement on the ancilla qubit yields
\begin{align}
&\mathrm{Im}[\bra{G[\bth^1]} P_0 P_1 U_\tau^\dagger[\bth^3] U_t^\dagger[\bth^2] P_2\, U_t[\bth^2] P_3\, U_\tau[\bth^3] P_4 P_5\ket{G[\bth^1]}] \nonumber \\
&=2p_{\ket{1}}-1\,,
\end{align}
where $p_{\ket{1}}$ is the probability for the ancilla to be measured in state $\ket{1}$.

\section{\label{sec:chi3app} Exact dynamics and spectrum of third-order susceptibility}

In this appendix, we derive analytical expressions for the third-order susceptibility in time and frequency domains. A more detailed discussion about the linear, second, and third-order susceptibilities of the quantum spin model eq~\ref{eq:Ham_spin} can be found in ref~\citenum{mootz2023twodimensional}.

By applying the completeness relation $\hat{I}=\sum_\nu\ket{\Psi_\nu}\bra{\Psi_\nu}$ with eigenstates $\ket{\Psi_\nu}$ of the quantum spin Hamiltonian $\h$,
eq~\ref{eq:chi3b} for $\alpha=\beta=\gamma=\delta=z$ can be written as
\begin{align}
\label{eq:chi3ex}
    &\chi^{(3)}_{zzzz}(t,\tau,0)=\frac{2}{N}\Theta(t)\Theta(\tau)\sum_{\mu\nu\lambda}F^z_{0,\mu}F^z_{\mu,\nu}F^z_{\nu,\lambda}F^z_{\lambda,0} \nonumber \\
    &\times \mathrm{Im}\left[\mathrm{e}^{\mathrm{i}\,E_0(t+\tau)-\mathrm{i}\,E_\mu t-\mathrm{i}\,E_\nu\tau} + \mathrm{e}^{\mathrm{i}\, E_\nu(t+\tau)-\mathrm{i}\,E_\lambda t-\mathrm{i}\,E_0\tau}\right. \nonumber \\
    &\qquad\quad\left. -2\,\mathrm{e}^{\mathrm{i}\,E_\mu(t+\tau)-\mathrm{i}\,E_\nu t-\mathrm{i}\,E_\lambda\tau}
    \right]\,.
\end{align}
Here, $E_\nu$ denotes the eigenenergies of $\h$ and $F^z_{j,k}\equiv\langle\Psi_j| \hat{S}^z|\Psi_k\rangle$ corresponds to the magnetic dipole matrix elements along the $z$-direction.   
Figure~\figref{fig6}{b} shows eq~\ref{eq:chi3ex} calculated for the two-site spin-1 model. The eigenenergies and eigenstates $E_\mu$ and $\Psi_\nu$ used to calculate $\chi^{(3)}_{zzzz}(t,\tau,0)$ were obtained using exact diagonalization. 

To transform $\chi^{(3)}_{zzzz}(t,\tau,0)$ to the 2D frequency space, the 2D Fourier transform
\begin{align}
	&\chi^{(3)}_{zzzz}(\omega_t,\omega_\tau,0)& \nonumber \\
 &=\int_{0}^\infty \mathrm{d}t\int_{0}^\infty \mathrm{d}\tau\,\chi^{(3)}_{zzzz}(t,\tau,0)\mathrm{e}^{\mathrm{i}\,(\omega_t+\mathrm{i}\,0^+) t}\mathrm{e}^{\mathrm{i}\,(\omega_\tau+\mathrm{i}\,0^+)\tau}
\end{align}
is applied, yielding~\cite{mootz2023twodimensional}
\begin{align}
	&\chi^{(3)}_{zzzz}(\omega_t,\omega_\tau,0)\nonumber \\
 &=\frac{\mathrm{i}}{N}\sum_{\mu,\nu,\lambda}S^{\mu,\nu,\lambda}_{zzzz}\left[L_{0,\mu}(\omega_t)L_{0,\nu}(\omega_\tau)-L_{\mu,\nu}(\omega_t)L_{0,\nu}(\omega_\tau)\right. \nonumber \\ &\qquad\qquad\qquad\left.
 -2 L_{\mu,\nu}(\omega_t)L_{\mu,\lambda}(\omega_\tau)+2 L_{\nu,\lambda}(\omega_t)L_{\mu,\lambda}(\omega_\tau)\right. \nonumber \\ &\qquad\qquad\qquad\left.+L_{\nu,\lambda}(\omega_t)L_{\nu,0}(\omega_\tau)-L_{\lambda,0}(\omega_t)L_{\nu,0}(\omega_\tau)\right]\,, 
 \label{eq:chi3w}
\end{align}
where 
\begin{align}	
S^{\mu,\nu,\lambda}_{zzzz}\equiv F^z_{0,\mu}F^z_{\mu,\nu}F^z_{\nu,\lambda}F^z_{\lambda,0}\,,
\end{align}
and
\begin{align}
	L_{\mu,\nu}(\omega)=\frac{1}{\omega+\mathrm{i}\,0^+ + E_\mu-E_\nu}
\end{align}
with $0^+$ denoting an infinitesimal positive quantity. According to eq~\ref{eq:chi3w}, the peaks in the 2D spectra of the third-order susceptibility in Figure~\figref{fig6}{e} and \figref{fig6}{f} emerge at energies determined by differences between eigenenergies $E_\mu-E_\nu$ along both the $\omega_t$- and $\omega_\tau$-axes. Consequently, these energy positions along $\omega_t$ and $\omega_\tau$ in $\chi^{(3)}_{zzzz}(\omega_t,\omega_\tau,0)$ describe two transitions between distinct eigenstates. The magnitude of the peaks in the 2D spectra is governed by $S^{\mu,\nu,\lambda}_{zzzz}$, along with the count of contributing transitions to the signals.

\begin{figure}[t!]
\begin{center}
		\includegraphics[scale=0.47]{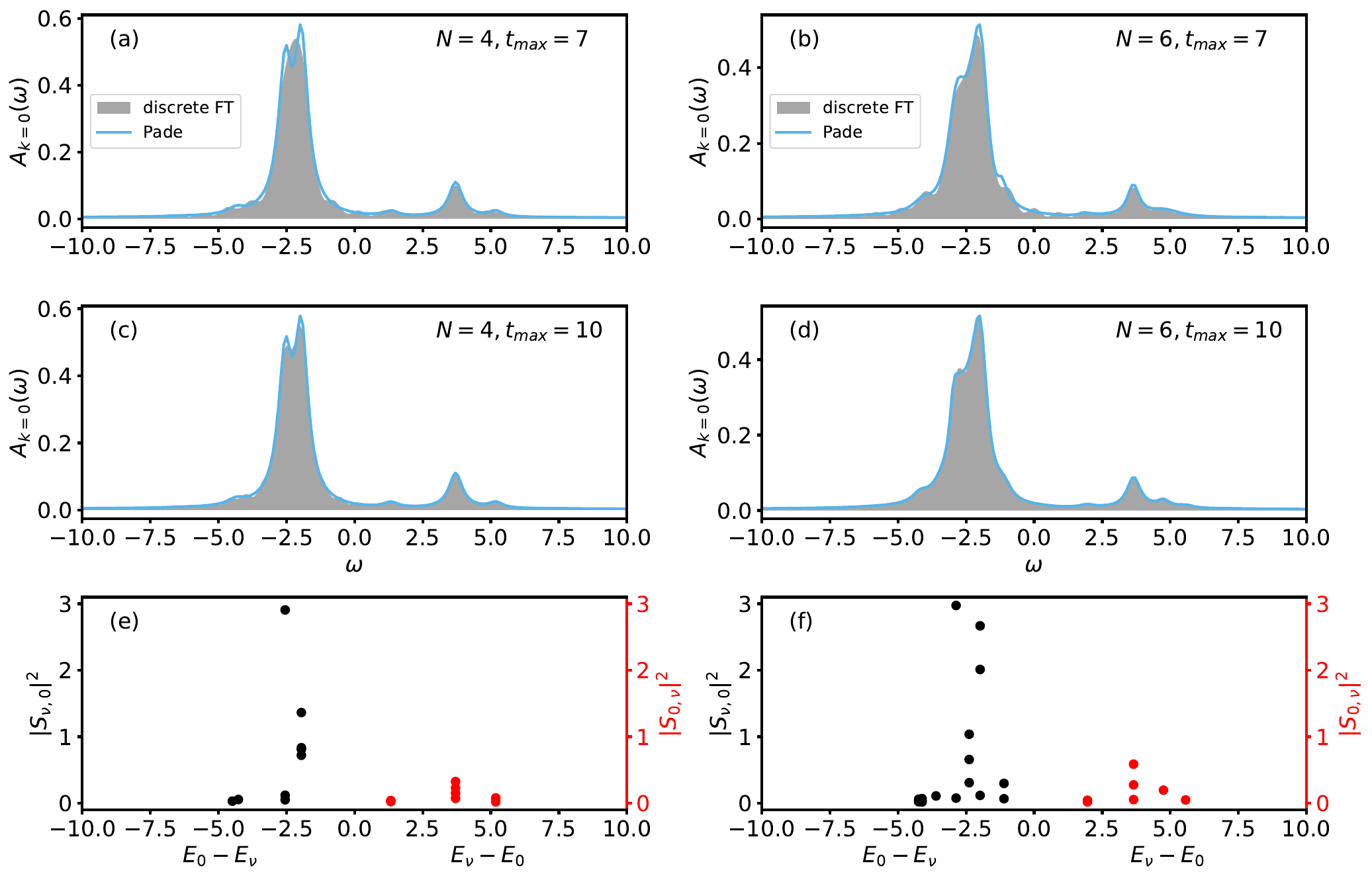}
		\caption{Spectral functions calculated with Pad\'e approximation vs. discrete Fourier transformation. (a), (b) Spectral function $A_{k=0}(\omega)$ obtained by Fourier transforming the Green's function dynamics presented in Figure~\figref{fig3}{a} for $N=4$ and \figref{fig3}{b} for $N=6$ and calculating eq~\ref{eq:Ak}. The spectral function obtained by using the discrete Fourier transformation eq~\ref{eq:pse} (shaded area) is show alongside the one calculated using the Pad\'e approximation (blue line). These results are obtained for a maximum simulation time of $t_\mathrm{max}=7$ and a damping factor of $\varepsilon=0.3$ in the Fourier transformation.  (c), (d) The corresponding results for $t_\mathrm{max}=10$. Compared to the spectral function obtained with Pad\'e approximation, $A_{k=0}(\omega)$ calculated with discrete Fourier transformation requires a larger maximum simulation time to achieve comparable accuracy. (e), (f) $|S_{\nu,0}|^2$ (black dots) and $|S_{0,\nu}|^2$ (red dots) as a function of the energy differences $E_0-E_\nu$ and $E_\nu - E_0$, respectively, for (e) $N=4$ and (f) $N=6$. Only the dominant transition amplitudes with $|S_{\nu,\mu}|^2 > 0.02$ are shown. The peaks in the spectral functions result from transitions between the ground state with energy $E_0$ to the excited states with energies $E_\nu$, and vice versa.}
		\label{fig7} 
\end{center}
\end{figure}

\section{\label{sec:FT} Spectral function calculated with Pad\'e approximation vs. discrete Fourier transformation}

In this appendix, we demonstrate that the Pad\'e approximation accelerates the convergence of the Fourier transform with respect to the required maximum simulation time compared to discrete Fourier transformation. Figure~\figref{fig7}{a} and \figref{fig7}{b} present the spectral function $A_{k=0}(\omega)$ obtained by calculating the Fourier transform of the real-time Green’s function shown in Figure~\figref{fig3}{a} for $N=4$ and Figure~\figref{fig3}{b} for $N=6$, and evaluating eq~\ref{eq:Ak}. The spectral function obtained using the discrete Fourier transformation eq~\ref{eq:pse} (shaded area) is compared with the one calculated using the Pad\'e approximation (blue line). These results are obtained for a maximum simulation time of $t_\mathrm{max}=7$ and a damping factor of $\varepsilon=0.3$ in the Fourier transformation. The corresponding results for $t_\mathrm{max}=10$ are shown in Figure~\figref{fig7}{c} and \figref{fig7}{d}. The spectral function obtained with the Pad\'e approximation using $t_\mathrm{max}=7$ resolves the dominant transitions between the ground state with energy $E_0$ to the excited states with energies $E_\nu$, and vice versa. This is seen by comparing the spectral functions in Figure~\figref{fig7}{a} and \figref{fig7}{b} with the corresponding transition amplitudes $|S_{\nu,0}|^2$ (black dots) and $|S_{0,\nu}|^2$ (red dots) plotted in Figure~\figref{fig7}{e} and \figref{fig7}{f}. In contrast, the spectral function obtained with discrete Fourier transformation using $t_\mathrm{max}=7$ does not resolve all dominant transitions in Figure~\figref{fig7}{e} and \figref{fig7}{f}. For example, the splitting of the main peak in Figure~\figref{fig7}{a} is not resolved in the spectral function calculated with the discrete Fourier transformation (shaded area). This demonstrates that the accurate calculation of the spectral function with the discrete Fourier transformation requires a larger $t_\mathrm{max}$. Specifically,  for a maximum simulation time of $t_\mathrm{max}=10$ (Figure~\figref{fig7}{c} and \figref{fig7}{d}), the spectral functions obtained with the discrete Fourier transformation show comparable accuracy to those calculated with the Pad\'e approximation.

\section{Prony approximation \label{sec:Prony}}

The Prony approximation is a powerful technique for decomposing a signal into a sum of damped exponentials. When applied to the Green's function $G(t)$, this method allows the identification of the complex frequencies (poles) and corresponding residues that characterize the system's spectral properties. Specifically, the Green's functions, sampled at $N_t$ discrete time points $t_n$, is expressed as a sum of $L$ complex exponentials:
\begin{align}
\label{eq:G_prony}
    G(t_n) = \sum_{i=1}^{L} c_i e^{-\gamma_i t_n}\,,
\end{align}
where $c_i$ are complex coefficients and $\gamma_i = \alpha_i + i\omega_i$ are the complex frequencies with damping factor $\alpha_i$ (related to broadening), and oscillation frequency $\omega_i$ (related to the energy levels). A linear prediction model is then constructed where each data point is predicted using a linear combination of the previous $L$ data points:
\begin{align}
    G(t_n) = -\sum_{i=1}^{L} a_i G(t_{n-i}) \quad \text{for } n \geq L\,.
\end{align}
The coefficients $a_i$ are determined by solving a linear system obtained from this model. The roots of the characteristic polynomial
\begin{align}
    p(z) = z^L + a_1 z^{L-1} + \dots + a_L
\end{align}
yields $z_i = e^{\gamma_i \Delta t}$ ($i=1,\dots,L$) where $\Delta t = t_{1} - t_{0}$ represents the uniform time step of the signal. Once the $z_i$ (or equivalently $\gamma_i$) are determined, the residues $c_i$ are calculated by solving a set of linear equations derived from the original signal data using the previously found $\gamma_i$. These residues represent the weights of each exponential component in the Green's function. The spectral function $A(\omega)$ is then constructed from the poles and residues. For each pole $\gamma_i$, a corresponding delta-like contribution is added to the spectral function:
\begin{align}
    A(\omega) =-\frac{1}{\pi}\mathrm{Im}\left[\sum_{j=1}^{L} \frac{c_j}{i(\omega - \omega_j) + \alpha_j}\right]\,.
\end{align}
This representation effectively captures the spectral features, with broadening determined by the real part $\alpha_j$ of the poles and the location of the peaks given by $\omega_j$. In the simulations, we include $L=100$ terms in the expansion of the Prony approximation (eq~\eqref{eq:G_prony})  which is sufficient to capture all dominant signals in the spectral function.

\section{Compressive sensing \label{sec:CS}}

Compressive sensing is a technique used to recover sparse signals from incomplete or noisy data. In the context of our analysis, the spectral function $A(\omega)$ is assumed to be sparse, meaning it can be represented as a sum of a few significant delta-like peaks. To apply compressive sensing, we start with the discretized time-domain Green's function $G(t_n)$, representing it as a linear combination of possible spectral components. This leads to a system of linear equations, where the goal is to find the spectral function $A(\omega)$ that best fits the observed data:
\begin{align}
G(t_n) = \int_{-\infty}^{\infty} A(\omega) e^{-i\omega t_n} d\omega.
\end{align}
This equation can be discretized over a grid of $N_\omega$ frequencies $\omega_m$ to yield the following linear system:
\begin{align}
G(t_n) \approx \sum_{m=1}^{N_\omega} A(\omega_m) e^{-i\omega_m t_n},
\end{align}
or in matrix form $\mathbf{G} = \mathbf{\Phi} \mathbf{A}$, where $\mathbf{G}$ is the vector of sampled values $G(t_n)$, $\mathbf{A}$ is the vector of spectral coefficients $A(\omega_m)$, and $\mathbf{\Phi}$ is the measurement matrix with elements $\Phi_{n,m} = e^{-i\omega_m t_n}$. The matrix $ \mathbf{\Phi}$ has dimensions $N_t \times N_\omega$. Using compressive sensing, we solve the underdetermined system $\mathbf{G} = \mathbf{\Phi} \mathbf{A}$ by enforcing sparsity on the solution vector $\mathbf{A} $. This is done by minimizing the $\ell_1$-norm of $\mathbf{A}$, subject to the constraint that the reconstructed Green's function matches the observed data within a certain tolerance. This problem can be formulated as:
\begin{align}
\mathbf{A}_\mathrm{min} = \text{argmin}_{\mathbf{A}} \left( \| \mathbf{G} - \mathbf{\Phi} \mathbf{A} \|_2^2 + \lambda \| \mathbf{A} \|_1 \right),
\end{align}
where $\lambda$ is a regularization parameter controlling the trade-off between the fidelity to the data (first term) and the sparsity of the solution (second term). The solution to this optimization problem is obtained using Python's LASSO (Least Absolute Shrinkage and Selection Operator) function, which efficiently enforces sparsity by shrinking small coefficients to zero while preserving the significant ones. The resulting spectral function $\mathbf{A}_\mathrm{min}(\omega)$ is highly accurate in reconstructing sharp features, such as delta peaks, even when the data contains noise. This method is particularly effective for cases with zero broadening, where it outperforms both the Pad\'e and Prony approximations as demonstrated in section~\ref{sec:GF_res}. In the calculation of the spectral functions, we use a regularization parameter of $\lambda=2\times 10^{-3}$, which results in an mean squared error of less than $0.3~\%$ between the actual Green's function in the time domain and the one predicted by the spectral function using LASSO. As a general trend, the number of observable peaks in the $A(\omega)$ increases with decreasing $\lambda$.

\begin{acknowledgement}
We acknowledge useful discussions with P. P. Orth and N. Gomes. We are grateful to the reviewers for many insightful suggestions, including additional signal processing techniques and an alternative way to measure response functions by leveraging mid-circuit measurements of the ancilla qubit.
This work was supported by the U.S. Department of Energy (DOE), Office of Science, Basic Energy Sciences, Materials Science and Engineering Division, including the grant of computer time at the National Energy Research Scientific Computing Center (NERSC) in Berkeley, California. The research was performed at the Ames National Laboratory, which is operated for the U.S. DOE by Iowa State University under Contract No. DE-AC02-07CH11358.
\end{acknowledgement}

\bibliography{ref}

\end{document}